\documentclass[preprint,nofootinbib,floatfix,a4paper,prd,superscriptaddress]{revtex4-1}   	
\usepackage{geometry}                		
\usepackage[colorlinks,citecolor=blue]{hyperref}
\usepackage{amsmath}
\usepackage{mathrsfs}
\usepackage{bbm}
\usepackage{amsfonts}
\usepackage{amssymb}
\usepackage{latexsym}
\usepackage{graphicx}
\usepackage[english]{babel}
\usepackage{multirow}
\usepackage{float}
\usepackage{url}
\usepackage{slashed}
\usepackage{xcolor} 
\usepackage{multirow}

\newcommand{\be}{\begin{equation}}
\newcommand{\ee}{\end{equation}}
\newcommand{\ba}{\begin{array}}
\newcommand{\ea}{\end{array}}
\newcommand{\bea}{\begin{eqnarray}}
\newcommand{\eea}{\end{eqnarray}}

\usepackage[utf8]{inputenc}
\usepackage[T1]{fontenc}
 \usepackage{CormorantGaramond}

\def  \bcen   {\begin{center}}
\def  \ecen   {\end{center}}
\def  \beq    {\begin{equation}}
\def  \eeq    {\end{equation}}

\def\la   {\lambda}

\def\lee { \left( }
\def\rii { \right) }

\def\to {\rightarrow}

\begin{document}

\title{Obliquely Scrutinizing a Hidden SM-like Gauge Model}

\author{Van Que Tran}
\email{vqtran@sjtu.edu.cn}
\affiliation{\small Tsung Dao Lee Institute $\&$ School of Physics and Astronomy, Shanghai Jiao Tong University, Shanghai 200240, China}
\affiliation{\small Faculty of Fundamental Sciences, PHENIKAA University, Yen Nghia, Ha Dong, Hanoi 12116, Vietnam}

\author{Thong T.~Q. Nguyen}
\email{ntqthonghep@gate.sinica.edu.tw}
\affiliation{\small Institute of Physics, Academia Sinica, Nangang, Taipei 11529, Taiwan}

\author{Tzu-Chiang Yuan}
\email{tcyuan@phys.sinica.edu.tw}
\affiliation{\small Institute of Physics, Academia Sinica, Nangang, Taipei 11529, Taiwan}

\date{\today}							

\begin{abstract}

In view of the recent high precision measurement of the Standard Model $W$ boson mass at the CDF II detector,
we compute the contributions to the oblique parameters $S$, $T$ and $U$ 
coming from the two additional Higgs doublets (one inert and one hidden) as well as the hidden neutral dark gauge bosons and 
extra heavy fermions in the gauged two-Higgs-doublet model (G2HDM). While the effects from the hidden Higgs doublet 
and new heavy fermions are found to be minuscule, the hidden gauge sector $SU(2)_H \times U(1)_X$ with gauge coupling strength 
$\gtrsim 10^{-2}$  and gauge boson mass $\gtrsim 100$ GeV 
can readily explain the $W$ boson mass anomaly but nevertheless  excluded by the dilepton  high-mass resonance 
searches at the Large Hadron Collider.
On the other hand, the new global fits to the oblique parameters due to new $W$ boson mass  measurement
can give discernible impacts on the  mass splitting and mixing angle for the inert Higgs doublet in G2HDM.
We also study the impact to the signal strength of diphoton mode of the 125 GeV Higgs boson $h \to \gamma \gamma$ and the 
detectability of the yet to observe process  $h \to Z \gamma$ at the High Luminosity Large Hadron Collider. 
Current constraints for the dark matter candidate $W^\prime$ including the dark matter relic density, dark matter direct detections and invisible Higgs decays 
are also taken into account in this study.

\end{abstract}

\maketitle

\section{Introduction}
\label{sec1}

Based on the data with an integrated luminosity of 8.8 fb$^{-1}$ collected by the CDF II detector between 2002 and 2011,
after over 10 years of dedicated analysis,
the CDF Collaboration at the Tevatron Collider has recently unveiled a high-precision direct measurement of the standard model (SM) $W$ boson mass.
The reported result is~\cite{CDF:2022hxs} 
\be
\label{mw-cdf}
m_W ({\rm CDF\; II}) = 80,433.5 \pm 9.4 \; {\rm MeV/}c^2 \; ,
\ee
which is $\sim 7 \sigma$ away from the SM prediction from electroweak (EW) global fits~\cite{deBlas:2021wap}
\be
m_W ({\rm SM - \; Global \, Fits}) = 80,359.1 \pm 5.2 \; {\rm MeV/}c^2 \; .
\ee
The CDF result (\ref{mw-cdf}) also represents a $\sim 3 \sigma$ deviation from other direct measurements from 
the more recent ATLAS~\cite{ATLAS:2017rzl} and LHCb~\cite{LHCb:2021bjt} experiments.
This immediately stirs a great deal of excitement in the field and triggers many subsequent studies.
Besides the impacts of this new measurement to the electroweak precision global fits~\cite{Lu:2022bgw,deBlas:2022hdk,Asadi:2022xiy}, 
it could entail new physics (NP) beyond the SM (BSM). 
We note that the deviation of the CDF $W$ boson mass measurement with the global fit only shifts
slightly from $7\sigma$ to 6$\sigma$  if the theoretical calculation tools used by CDF is updated by a more recent version~\cite{Isaacson:2022rts}.

While combined result of the measurements from  LEP, Tevatron and ATLAS are still lacking, pending on evaluation of uncertainty correlations~\cite{CDF:2022hxs}, 
and the new CDF result of $m_W$ is needed to be independently confirmed, 
BSM enthusiast has already offering various NP interpretations of the new CDF result.
See for example, Refs.~\cite{Du:2022fqv,Zeng:2022lkk,Cheng:2022aau} for extra $U(1)$ implications, 
\cite{Fan:2022dck,Sakurai:2022hwh,Cheng:2022jyi,Babu:2022pdn,Heo:2022dey,Ahn:2022xeq,Arcadi:2022dmt,Lee:2022gyf,Abouabid:2022lpg,Du:2022brr,Mondal:2022xdy,Kanemura:2022ahw,Chen:2022ocr,Bahl:2022gqg} for extended Higgs sectors, \cite{Fan:2022yly,Bagnaschi:2022whn} for SMEFT, 
\cite{Cheung:2022zsb,Lee:2022nqz} for lepto-quark, \cite{Wang:2022dte,Kim:2022zhj} for dark matter (DM) models, 
\cite{Yang:2022gvz,Tang:2022pxh,Zheng:2022irz,Ghoshal:2022vzo} for low energy supersymmetry, 
and \cite{Evans:2022dgq,Senjanovic:2022zwy,Barger:2022wih} for grand unification, {\it etc.} 
Another interesting point to support the BSM physics has been pointed out in~\cite{Athron:2022qpo} is that
the hadronic uncertainties in the fine structure constant and hadronic vacuum polarization that 
affects the SM $W$ mass and muon anomalous magnetic dipole moment respectively are anti-correlated with each other.

In this work, we study the impact of the new CDF result to the extra particle mass spectra in the gauged two-Higgs-doublet model (G2HDM) first proposed in~\cite{Huang:2015wts} and explored further in~\cite{Huang:2015rkj,Arhrib:2018sbz,Huang:2019obt,Huang:2017bto,Chen:2018wjl,Chen:2019pnt,Dirgantara:2020lqy,Ramos:2021txu,Ramos:2021omo}.
G2HDM is a gauged DM model based on the extended electroweak gauge group $G_{\rm G2HDM} = SU(2)_L \times U(1)_Y \times SU(2)_H \times U(1)_X$.
The extra gauge group $SU(2)_H \times U(1)_X$ represents the dark gauge sector interacting feebly with the visible SM sector, in the sense that the new 
gauge couplings $g_H$ and $g_X$ are much smaller than the SM electroweak gauge couplings $g$ and $g^\prime$,
as suggested by our recent detailed studies~\cite{Ramos:2021txu,Ramos:2021omo}.
Additional Higgses and heavy fermions must be included in G2HDM to make it phenomenologically viable and free from gauge and gravitational anomalies. 
The crucial idea of G2HDM is that the usual two Higgs doublets ($H_1$ and $H_2$) in general two-Higgs-doublet model (2HDM) is lumped together in
a 2 dimensional spinor representation $H= (H_1  \, H_2)^{\rm T}$ of a hidden local $SU(2)_H$ gauge group.
One distinctive feature of the model is that there is no need to impose an {\it ad hoc} discrete $Z_2$ symmetry to stabilize the DM. 
There is a hidden $h$-parity in the model~\cite{Chen:2019pnt}, 
admitted readily once one writes down all possible gauge invariant and renormalizable interactions, that will guarantee the lightest $h$-parity odd particle to be the DM candidate,
provided that it is not broken spontaneously. 
Another interesting feature is that the new gauge bosons $W^{\prime \, (p,m)} \equiv (W^\prime_1 \mp i W^\prime_2)/\sqrt 2$ 
are electrically neutral and don't mix with the SM $W^\pm$ bosons. They are also $h$-parity odd and can be the DM candidate. 
A scenario of sub-GeV low mass $W^{\prime \, (p,m) }$ as DM was studied in~\cite{Ramos:2021txu,Ramos:2021omo}. 
In this work, we will turn our attention to the scenario of $W^{\prime \, (p,m) }$ as DM candidate with a wider mass range.

In the next section \ref{sec2}, we will briefly review the particle content in the minimal G2HDM~\cite{Ramos:2021txu,Ramos:2021omo}, their masses 
and interactions that are relevant to our study. In section \ref{sec3}, we remind ourselves by reviewing the Peskin-Takeuchi oblique parameters~\cite{Peskin:1991sw} 
that not only entered in the art of global fit analysis which correlates all electroweak observables but also provided important constraints on NP models.
In section \ref{sec4}, we compute the new contributions to the oblique parameters from the 
mixings of the neutral gauge bosons (section \ref{subsec4a}),
extended Higgs sector (sections \ref{subsec4b} and \ref{subsec4c})
and extra heavy fermions (section \ref{subsec4d}) in G2HDM. 
In section \ref{sec5}, we present and discuss our numerical results. In addition, we will take the opportunity in this section to explore the detectability 
of the process $h \to Z \gamma$ in the High Luminosity Large Hadron Collider (HL-LHC).
Conclusions are given in section \ref{sec6}. We reserve an appendix for the analytical expressions of the  one-loop 
amplitudes of the two processes $h_i \to \gamma \gamma$ ($i=1,2$) and $h_i \to Z_j \gamma$ ($i=1,2;j=1,2,3$) in G2HDM,
where $h_1$ and $Z_1$ are identified as the SM 125 GeV Higgs scalar ($h$) and 91 GeV $Z$ vector boson respectively.

\section{Model Setup}
\label{sec2}

In this section, we will briefly review the minimal G2HDM. The original model was introduced in Ref.~\cite{Huang:2015wts},
and various refinements~\cite{Arhrib:2018sbz,Huang:2019obt,Chen:2019pnt} and 
collider implications~\cite{Huang:2015rkj,Chen:2018wjl,Huang:2017bto} were pursued subsequently with the same particle content as 
the original model. 
As advocated recently in~\cite{Ramos:2021omo,Ramos:2021txu}, we will drop the triplet field $\Delta_H$ of the extra $SU(2)_H$  
which can vastly simplify the scalar potential by getting rid of 6 parameters. We will refer this as the minimal G2HDM in what follows.
The quantum numbers of the matter particles in G2HDM under $SU(3)_C \times SU(2)_L \times SU(2)_H \times U(1)_Y \times U(1)_X$ 
are~\footnote{The last two entries in the tuples are the hypercharge and $X$ charge of the two $U(1)$ factors. 
Note that fields with $Q_X = \pm 1$ in our earlier 
works~\cite{Huang:2015wts,Arhrib:2018sbz,Huang:2019obt,Chen:2018wjl,Huang:2017bto,Chen:2019pnt,Ramos:2021omo,Ramos:2021txu}
have been changed to $\pm 1/2$. The anomaly cancellations of the model remain intact with these changes.}\\
\noindent
\underline{Scalars}:
$$H=\left( H_1 \;\; H_2 \right)^{\rm T} \sim  \left( {\bf 1}, {\bf 2}, {\bf 2}, \frac{1}{2}, { \frac{1}{2} }\right) \; , \; 
\Phi_H=\left( \Phi_1 \;\; \Phi_2 \right)^{\rm T} \sim \left( {\bf 1}, {\bf 1}, {\bf 2}, 0, { \frac{1}{2} } \right) \; ; $$

\noindent
\underline{Spin 1/2 Fermions}:

\underline{Quarks}
$$Q_L=\left( u_L \;\; d_L \right)^{\rm T} \sim \left(  {\bf 3}, {\bf 2}, {\bf 1}, \frac{1}{6}, 0 \right) \; , \; 
U_R=\left( u_R \;\; u^H_R \right)^{\rm T} \sim  \left( {\bf 3}, {\bf 1}, {\bf 2}, \frac{2}{3},  \frac{1}{2}  \right) \; , $$
$$D_R=\left( d^H_R \;\; d_R \right)^{\rm T} \sim \left( {\bf 3}, {\bf 1}, {\bf 2},  -\frac{1}{3}, - \frac{1}{2}  \right) \; ; $$
$$u_L^H \sim \left(  {\bf 3}, {\bf 1}, {\bf 1},  \frac{2}{3}, 0 \right) \; , \; d_L^H \sim \left(  {\bf 3}, {\bf 1}, {\bf 1}, -\frac{1}{3}, 0 \right) \; ; $$

\underline{Leptons}
$$L_L=\left( \nu_L \;\; e_L \right)^{\rm T} \sim \left( {\bf 1}, {\bf 2}, {\bf 1},  -\frac{1}{2}, 0 \right) \; , \; 
N_R=\left( \nu_R \;\; \nu^H_R \right)^{\rm T} \sim \left( {\bf 1}, {\bf 1}, {\bf 2},  0,  \frac{1}{2}  \right)  \; , $$
$$E_R=\left( e^H_R \;\; e_R \right)^{\rm T} \sim \left( {\bf 1}, {\bf 1}, {\bf 2},  -1, - \frac{1}{2}  \right) \; ; $$
$$\nu_L^H \sim \left( {\bf 1}, {\bf 1}, {\bf 1},  0, 0 \right) \; , \; e_L^H \sim \left( {\bf 1}, {\bf 1}, {\bf 1},  -1, 0 \right) \; .$$

The most general renormalizable Higgs potential which is invariant under both $SU(2)_L\times U(1)_Y$ and  $SU(2)_H \times  U(1)_X$  
can be written down as follows
\begin{align}\label{eq:V}
V = {}& - \mu^2_H   \left(H^{\alpha i}  H_{\alpha i} \right) - \mu^2_{\Phi}   \Phi_H^\dag \Phi_H  
+  \lambda_H \left(H^{\alpha i}  H_{\alpha i} \right)^2  + \la_\Phi \lee \Phi_H^\dag \Phi_H  \rii^2 \nonumber \\
{}&+ \frac{1}{2} \lambda'_H \epsilon_{\alpha \beta} \epsilon^{\gamma \delta}
\left(H^{ \alpha i}  H_{\gamma  i} \right)  \left(H^{ \beta j}  H_{\delta j} \right)  \\
{}&  
+\lambda_{H\Phi} \lee H^\dag H  \rii  \lee \Phi_H^\dag \Phi_H \rii  
 + \lambda^\prime_{H\Phi} \lee H^\dag \Phi_H  \rii  \lee \Phi_H^\dag H \rii,  \nonumber 
\end{align}
where  ($i$, $j$)  and ($\alpha$, $\beta$, $\gamma$, $\delta$) refer to the $SU(2)_L$ and $SU(2)_H$ indices respectively, 
all of which run from one to two, and $H^{\alpha i} = H^*_{\alpha i}$.

To study spontaneous symmetry breaking (SSB) in the model, we parameterize the Higgs fields according to standard lore
\begin{eqnarray}
\label{eq:scalarfields}
H_1 = 
\begin{pmatrix}
G^+ \\ \frac{ v + h_{\rm SM}}{\sqrt 2} + i \frac{G^0}{\sqrt 2}
\end{pmatrix}
, \,
H_2 = 
\begin{pmatrix}
 H^+  \\  H_2^0 
\end{pmatrix}
, \,
\Phi_H = 
\begin{pmatrix}
G_H^p  \\ \frac{ v_\Phi + \phi_H}{\sqrt 2} + i \frac{G_H^0}{\sqrt 2}
\end{pmatrix}
\; \; \;
\end{eqnarray}
where $v$ and $v_\Phi$ are the only non-vanishing vacuum expectation values (VEVs)
in $H_1$ and $\Phi_{H}$ fields respectively.  $v = 246$ GeV is the SM VEV.

The relevant interaction Lagrangian for the computation of the one-loop oblique parameters in G2HDM is 
\beq
\label{Lint}
\mathcal L_{\rm int}  = \mathcal L_{\rm int \, 1} + \mathcal L_{\rm int \, 2}  \; , 
\eeq
where
\bea 
\label{Lint1}
\mathcal L_{\rm int \, 1} & \supset &
- \frac{1}{2} \left( \partial_\mu h_{\rm SM} \right) \left[ \left( g W_3^\mu - g^\prime B^\mu \right) G^0 + i g \left( G^+ W^{- \mu} - G^- W^{+ \mu} \right) \right] \nonumber \\
& + & \frac{1}{2} \left( h_{\rm SM} + v \right)  \left\{  \left(  \partial_\mu G^0 \right) \left( g W_3^\mu - g^\prime B^\mu \right) 
+ i g \left[ \left( \partial_\mu G^+ \right) W^{- \mu} - \left( \partial_\mu G^- \right) W^{+\mu} \right] \right\}
 \nonumber \\
& + & \frac{i}{2} \left( g W_3^\mu - g^\prime B^\mu \right) \left[ \left( \partial_\mu H^{0 \, *}_2 \right) H_2^0 -  \left( \partial_\mu H^{0}_2 \right) H_2^{0 \, *}  \right]  \\
& + & \frac{i}{2} \left( g W_3^\mu + g^\prime B^\mu \right) \left[  \left( \partial_\mu H^{+} \right) H^- -  \left( \partial_\mu H^{-} \right) H^{+} \right]  \nonumber \\
& + & i \frac{g}{\sqrt 2}   \left\{ W^{-\mu} \left[ \left( \partial_\mu H^{+} \right) H_2^{0 \, *} -  \left( \partial_\mu H^{0 \,*}_2 \right) H^{+} \right]
- W^{+\mu}  \left[ \left( \partial_\mu H^{-} \right) H_2^{0} -  \left( \partial_\mu H^{0}_2 \right) H^{-}  \right] \right\}  \nonumber  \\
& + & \cdots \;,  \nonumber 
\eea
and 
\bea
\label{Lint2}
\mathcal L_{\rm int \, 2} & \supset &
 \frac{1}{8} \left[ g^2  \left( W^\mu_3 W_{3 \mu} + 2 \, W^{+ \mu} W^{-}_{\mu} \right) + g^{\prime \, 2} B^\mu B_\mu \right]
 \left[ \left( h_{\rm SM} + v \right)^2 + 2 \left( H_2^{0\, *} H_2^0 + H^+ H^- \right) \right]   \nonumber \\
& - & \frac{1}{4} g g^\prime  W^\mu_3 B_\mu \left[ \left( h_{\rm SM} + v \right)^2 + 2 \left( H_2^{0\, *} H_2^0 - H^+ H^- \right) \right]  + \cdots  \; .
\eea
Note that the $\cdots$ in (\ref{Lint1}) and (\ref{Lint2}) indicate terms of first and second order in $g_H$ and $g_X$ have been ignored under our approximations.
Their effects will be taking into account at the tree level via the mass mixings in the neutral gauge bosons in the model 
as will be explained further in section \ref{subsec4a}.

In G2HDM, the SM $Z_{\rm SM}$ and $A$ fields are defined as usual
\beq
\begin{pmatrix}
W_3^ \mu \\
B^\mu 
\end{pmatrix} 
= 
\begin{pmatrix}
c_W & s_W  \\
- s_W &  c_W 
\end{pmatrix}
\begin{pmatrix}
 Z_{\rm SM}^\mu \\
A^\mu 
\end{pmatrix} \; ,
\eeq
where 
\be
s_W \equiv \sin \theta_W = \frac{ g^\prime } {\sqrt{g^2 + g'^2 }} \;, \;\;\;\;\;  c_W \equiv \cos \theta_W = \frac{ g } {\sqrt{g^2 + g'^2 }} \; ,
\ee
and the electric charge $e$ is given by  
 \be 
 e = \frac{ g g'} {\sqrt{g^2 + g'^2 }} \;  \;\;\; {\rm and} \;\;\;\;  \alpha = \frac{e^2}{4 \pi}  \; . 
\ee
In G2HDM, the SM $W$ boson does not mix with $W^\prime$ and its mass is the same as in SM: $m_W = g v/2$.
However in general the SM $Z_{\rm SM}$ will mix further with the gauge field $W^{\prime}_3$ associated with the third generator of $SU(2)_H$ 
and the $U(1)_X$ gauge field $X$ via the following mass matrix:
\be
\mathcal M_Z^2 =  
\begin{pmatrix}
m^2_{Z} & - \frac{1}{2} g_H v m_{Z} & - { \frac{1}{2} } g_X v m_{Z} \\
 - \frac{1}{2} g_H v m_{Z} & m^2_{W^\prime} & { \frac{1}{4} } g_H g_X \left( v^2 - v^2_\Phi \right) \\
- {\frac{1}{2} } g_X v m_{Z} &   { \frac{1 }{4} } g_H g_X \left( v^2 - v^2_\Phi \right) & {\frac{1}{4} } g_X^2 \left( v^2 + v_\Phi^2 \right) + M_X^2
\end{pmatrix} \; ,
\label{MsqZs}
\ee
where 
\bea
 \label{mzsm}
 m_{Z} & = &  \frac{1 }{2}  v \sqrt{ g^2 + g^{\prime \, 2} }\; , \\
 \label{mwprime}
 m_{W^\prime} & = & \frac{1}{2} g_H \sqrt{ v^2 + v_\Phi^2 } \; ,
\eea
and $M_X$ is the Stueckelberg mass for the $U(1)_X$. 

The real and symmetric mass matrix $\mathcal M_Z^2 $ in (\ref{MsqZs}) can be diagonalized by a 3 by 3 orthogonal matrix ${\mathcal O}^G$,
{\it i.e.} $(\mathcal O^{G})^{\rm T} \mathcal M_Z^2 \mathcal O^G = {\rm Diag}(m^2_{Z_1}, m^2_{Z_2},m^2_{Z_3})$, where $m_{Z_i}$ is 
the mass of the physical fields $Z_i$ for $ i=1,2,3$.
We will identify $Z_1$ to be the neutral gauge boson resonance with a mass $m_{Z_1} = 91.1876$  GeV observed at LEP~~\cite{Zyla:2020zbs}. 
The lighter/heavier of the other two states is the dark photon ($\gamma^\prime$)/dark $Z$ ($Z^\prime$). These neutral gauge bosons are $h$-parity even in the model,
despite the adjective `dark' are used for the other two states. 
The DM candidate considered in this work is $W^{\prime \, (p,m)}$, which is electrically neutral but carries one unit of dark charge and
chosen to be the lightest $h$-parity odd particle in the parameter space.

In G2HDM there are mixings effects of the two doublets $H_1$ and $H_2$ with the hidden doublet $\Phi_H$.
The neutral components $h_{\rm SM}$ and $\phi_H$ in $H_1$ and $\Phi_H$ respectively are both $h$-parity even.
They mix to form two physical Higgs fields $h_1$ and $h_2$
\be
\left(
\begin{matrix}
h_{\rm SM} \\
\phi_H
\end{matrix}
\right)
= 
{\mathcal O}^S \cdot 
\left(
\begin{matrix}
h_1 \\
h_2
\end{matrix}
\right)
=
\left( 
\begin{matrix}
 \cos \theta_1  &  \sin \theta_1 \\
- \sin \theta_1  &  \cos \theta_1 
\end{matrix}
\right) \cdot 
\left(
\begin{matrix}
h_1 \\
h_2
\end{matrix}
\right)
\; .
\ee
The mixing angle $\theta_1$ is given by
\be
\tan 2 \theta_1 = \frac{\lambda_{H\Phi} v v_\Phi }{ \lambda_\Phi v^2_\Phi - \lambda_H v^2 } \; .
\ee
The masses of $h_1$ and $h_2$ are given by
\be
m_{h_1,h_2}^2 = \lambda_H v^2 + \lambda_\Phi v_\Phi^2 \mp 
\sqrt{\lambda_H^2 v^4 + \lambda_\Phi^2 v_\Phi^4 + \left( \lambda^2_{H\Phi}  - 2 \lambda_H \lambda_\Phi \right) v^2 v_\Phi^2 } \; .
\ee
The lighter state $h_1 \equiv h$ is identified as the observed Higgs boson at the LHC. 
Currently the most precise measurement of the Higgs boson mass is $m_{h_1} = 125.38 \pm 0.14$ GeV~\cite{CMS:2020xrn}.

The complex fields $H_2^{0 \, *}$ and $G^p_H$ in $H_2$ and $\Phi_H$ respectively are both $h$-parity odd. They
mix to form a physical dark Higgs $D^*$ and a unphysical Goldstone field $\tilde G^*$ absorbed by the $W^{\prime \, p}$
\bea
\label{H20field}
\begin{pmatrix}
G^m_H \\
H_2^{0} 
\end{pmatrix}
= \mathcal O^D \cdot
\begin{pmatrix}
 \tilde G \\
 D 
 \end{pmatrix}
=
\begin{pmatrix}
 \cos \theta_2 & \sin \theta_2  \\
- \sin \theta_2  & \cos \theta_2 
\end{pmatrix} 
\cdot
\begin{pmatrix}
 \tilde G \\
 D 
 \end{pmatrix}
 \; .
\eea
The mixing angle $\theta_2$ satisfies
\be
\label{theta2}
\tan 2 \theta_2 = \frac{2 v v_\Phi}{v^2_\Phi - v^2} \; ,
\ee
and the mass of $D$ is 
\be
m_D^2 = \frac{1}{2} \lambda^\prime_{H\Phi} \left( v^2 + v_\Phi^2 \right) \; .
\ee
In the Feynman-'t Hooft gauge the Goldstone field $\tilde G^*$ $(\tilde G )$ has the same mass as the $W^{\prime \, p}$ ($W^{\prime \, m}$) which is given by 
(\ref{mwprime}).
Finally the charged Higgs $H^\pm$ is also $h$-parity odd and has a mass
\be
m^2_{H^\pm} = \frac{1}{2} \left( \lambda^\prime_{H\Phi} v^2_\Phi - \lambda^\prime_H v^2 \right) \; .
\ee

One can do the inversion to express the fundamental parameters in the scalar potential in terms of the particle masses~\cite{Ramos:2021txu,Ramos:2021omo}:
\bea
v_\Phi & = & 
\begin{cases}
\begin{matrix}
v  \cot \theta_2 \, , &&  {\rm for} \; \theta_2 > 0 \; , \\
- v \tan \theta_2 \, , &&  {\rm for} \; \theta_2 \leq 0 \; , 
\end{matrix}
\end{cases} 
\\
\lambda_H & = & \frac{1}{2 v^2} \left( m^2_{h_1} \cos^2 \theta_1 + m^2_{h_2} \sin^2 \theta_1 \right) \; , \\
\lambda_\Phi & = & \frac{1}{2 v_\Phi^2} \left( m^2_{h_1} \sin^2 \theta_1 + m^2_{h_2} \cos^2 \theta_1 \right) \; , \\
\lambda_{H\Phi} & = & \frac{1}{2 v v_\Phi} \left( m^2_{h_2} - m^2_{h_1} \right) \sin \left( 2 \theta_1 \right) \; , \\
\lambda^\prime_{H\Phi} & = & \frac{ 2  m_D^2 }{v^2 + v_\Phi^2} \; , \\
\lambda^\prime_{H} & = & \frac{2}{v^2} \left( \frac{m_D^2 v_\Phi^2 }{ v^2 + v_\Phi^2 } - m^2_{H^\pm}\right) \; .
\eea
From (\ref{mwprime}), we also have
\beq
g_H  =   \frac{ 2 m_{W'} } { \sqrt{v^2 + v^2_{\Phi} }}  \;. 
\eeq
Thus one can use  $m_{h_2}$, $m_{W^\prime}$, $m_D$, $m_{H^\pm}$, $\theta_1$ and $\theta_2$ as input in our numerical scan. 

The Feynman rules can be straightforwardly derived by rewriting the two Lagrangians (\ref{Lint1}) and (\ref{Lint2}) using the above physical fields.
The $h$-parity odd particles in G2HDM are $W^{\prime \, (p,m)}$, $D^{(*)}$, ${\tilde G}^{(*)}$, $H^\pm$, and all new heavy fermions $f^H$.
Among them, $W^{\prime \, (p,m)}$, $D^{(*)}$, and $\nu^H$ are electrically neutral and hence any one of them can be a DM candidate.
The DM phenomenology of complex scalar $D^{(*)}$ was studied in detail in~\cite{Chen:2019pnt,Dirgantara:2020lqy} and for low mass 
$W^{\prime \, (p,m)}$ as DM, see~\cite{Ramos:2021omo,Ramos:2021txu}.
For further details of G2HDM, we refer our readers to the earlier works~\cite{Huang:2019obt,Arhrib:2018sbz}. 

\section{Oblique Parameters and $W$ Boson Mass Shift}
\label{sec3}

As is well known, the oblique parameters $S$, $T$, and $U$~\cite{Peskin:1991sw} represent the most important electroweak radiative corrections since they are defined by
the transverse pieces of the vacuum polarization tensors of the SM vector gauge bosons. They are process independent whereas 
the other vertex and box corrections are necessarily attached to the particles in the initial and final states in the elementary processes in high precision experiments. 

The vacuum polarization tensor $i \Pi_{IJ}^{\mu\nu}(q)$ involving the SM gauge bosons $I$ and $J$ has the following decomposition
\be
i \Pi_{IJ}^{\mu\nu}(q) = i \left( \Pi_{IJ} (q^2) g^{\mu\nu} - \Delta_{IJ} (q^2) q^\mu q^\nu \right) \; .  
\ee
The form factor $\Delta_{IJ}(q^2)$ needs no concern to us since at  high energy experiments like LEP I and II where electroweak precision 
measurements were carried out, $q^\mu$ will dot into the helicity spinors of light leptons 
and will give vanishing results. The vacuum polarization amplitude $\Pi_{IJ} (q^2)$ has the following expansion
\bea
\Pi_{\gamma\gamma} (q^2) & =  & q^2 \Pi^\prime_{\gamma\gamma}(0) + \cdots \; , \\
\Pi_{Z\gamma} (q^2) & =  & q^2 \Pi^\prime_{Z\gamma}(0) + \cdots \; , \\
\Pi_{ZZ} (q^2) & = & \Pi_{ZZ}(0) + q^2 \Pi^\prime_{ZZ}(0) + \cdots \; , \\
\Pi_{WW} (q^2) & = & \Pi_{WW}(0) + q^2 \Pi^\prime_{WW}(0) + \cdots \; .
\eea

The oblique parameters $S$, $T$ and $U$ are defined with an overall factor of $\hat \alpha=\hat e^2/4\pi$ extracted out in front as~\cite{Peskin:1991sw}
\bea
\label{S}
\hat \alpha S & = & 4 \hat s_W^2 \hat c_W^2 \left[ \Pi^\prime_{ZZ}(0)  - \frac{\hat c_W^2 - \hat s_W^2}{\hat s_W \hat c_W} \Pi^\prime_{Z\gamma}(0)  - \Pi^\prime_{\gamma\gamma}(0) \right]  \; , \\
\label{T}
\hat \alpha T & = & \frac{\Pi_{WW}(0)}{m_W^2} - \frac{\Pi_{ZZ}(0)}{m_Z^2} \; , \\
\label{U}
\hat \alpha U & = & 4 \hat s_W^2  \left[ \Pi^\prime_{WW}(0)  - \hat c_W^2 \Pi^\prime_{ZZ}(0)  - 2 \hat s_W \hat c_W \Pi^\prime_{Z\gamma}(0)  - \hat s_W^2 \Pi^\prime_{\gamma\gamma}(0) \right] \; .
\eea
The $W$ boson mass shift can be related to the oblique parameters according to~\cite{Peskin:1991sw}
\bea
\label{wmassshift}
\frac{ \Delta m_W^2 }{m_Z^2}  & = & \hat \alpha \frac{\hat c_W^2 }{\hat c_W^2 - \hat s_W^2} \left[ -\frac{S}{2} + \hat c_W^2 T 
+ \frac{\hat c_W^2 - \hat s_W^2}{4 \hat s_W^2} U  \right] \; .
\eea
Here the hat quantities $\hat c_W$, $\hat s_W$ and $\hat \alpha$ are understood to be evaluated at the $Z$ pole. 
In order to compare with the experimental value $\Delta m_W ({\rm CDF \, II}) \approx$ 75 MeV  from (\ref{mw-cdf}), 
we use $\Delta m_W \approx  \left( \Delta m_W^2 + m_W^2 \right)^{1/2} - m_W$ with $m_W$ given 
by the SM expression $m_W = g v/2$ and $\Delta m_W^2$ by (\ref{wmassshift}).

\section{New Contributions to the Oblique Parameters in G2HDM}
\label{sec4}

In this section, we compute the contributions to the oblique parameters from all the new particles introduced in G2HDM.
First, we will handle the tree level contributions to the oblique parameters coming from the mass mixing matrix for the three 
massive neutral gauge bosons given in (\ref{MsqZs}).

\subsection{Contributions from the tree level mixings of $Z_{\rm SM}$, $W^\prime_3$ and $X$} 
\label{subsec4a}

\begin{figure}[tb] 
\includegraphics[width=12cm,height=1.0cm]{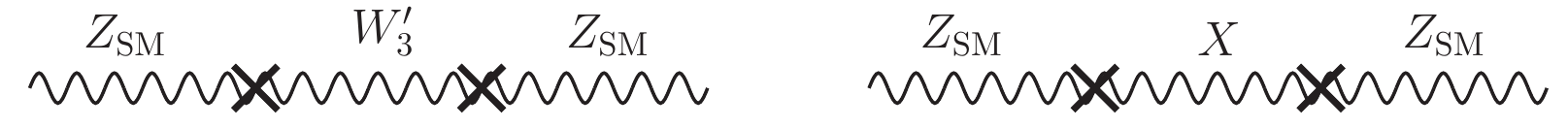}  
   \caption{Tree level diagrams (to second orders of $g_H$ and $g_X$) that contribute to the oblique parameters 
   from the mass mixings (\ref{MsqZs}) between the $Z_{\rm SM}$, $W^\prime_3$ and $X$ in G2HDM.}
   \label{STU-NGBMixings}
\end{figure}

The kinetic and mass mixings of an extra $U(1)$ boson with the SM $B$ and $W^3$ gauge fields
violate the custodial symmetry in the SM at the tree level and can give rise to the effective shift 
of the oblique parameters~\cite{Holdom:1990xp} (see also \cite{Cheng:2022aau}). 
The neutral gauge boson mass mixings (\ref{MsqZs}) of $Z_{\rm SM}$, $W^\prime_3$ and $X$ in G2HDM also 
violate the custodial symmetry in the SM and can give rise to a non-vanishing vacuum polarization amplitude $\Pi_{ZZ}(q)$ 
at the tree level. The other 3 vacuum polarization amplitudes vanish at tree level in the model.
For $g_H, g_X \ll g, g^\prime$, from the Feynman diagram in Fig.~\ref{STU-NGBMixings}, one obtains
\be
\label{PZZ-Wp3-X}
\Pi_{ZZ} (q^2) \approx \frac{1}{4} v^2 m_Z^2 \left( \frac{g_H^2}{q^2 - m_{W^\prime}^2} + \frac{g_X^2}{q^2 - m_X^2}  \right) \; ,
\ee
where $m^2_{W^\prime}$ is given by (\ref{mwprime}) and $m_X^2$ is the 33 entry of  (\ref{MsqZs}), namely
\be
\label{mXMX}
m_X^2 = \left( \mathcal M_Z^2 \right)_{33} = \frac{1}{4} g_X^2 \left( v^2 + v_\Phi^2 \right) + M_X^2 \; .
\ee
The $Z$ boson propagator is modified as
\be
\label{modifiedZprop}
i \Delta_Z^{\mu\nu} (q) =  - i \left( \frac{ g^{\mu\nu} } {q^2 - m_Z^2 - \Pi_{ZZ}(q^2) } - q^{\mu} q^{\nu} \; {\rm term} \right) \; .
\ee
 This modified propagator implies a mass correction $\delta m^2_Z$ and a wave function renormalization constant $\mathcal Z$ for the $Z$ boson, {\it viz.}
\bea
\delta m_Z^2 & \approx & \Pi_{ZZ}(m_Z^2)  - m_Z^2 \Pi^\prime_{ZZ} (m_Z^2)  \; , \\
\sqrt \mathcal Z & \approx & 1 + \frac{1}{2} \Pi_{ZZ}^\prime (m_Z^2) \; .
\eea
The mass shift for the physical $Z$ field is then given by $\Delta m_Z^2 \approx \delta m_Z^2 + m_Z^2 ( \mathcal Z - 1) \approx \Pi_{ZZ}(m_Z^2)$~\footnote{
$\Delta m_Z$ can then be computed as $\Delta m_Z \approx \left( \Delta m_Z^2 + m_Z^2 \right)^{1/2} - m_Z \approx \left( \Pi_{ZZ}(m_Z^2) + m_Z^2 \right)^{1/2} - m_Z$ with $m_Z$ given by the SM expression (\ref{mzsm}) and $\Pi_{ZZ}(m_Z^2)$ by (\ref{PZZ-Wp3-X}).}.
One can then deduce the effective oblique parameters by using the EFT approach~\cite{Burgess:1993vc}. 
In terms of the effective parameters $C$ and $z$ defined by Eq.~(1) in~\cite{Burgess:1993vc}, we have $C = - \Pi^\prime_{ZZ}(m_Z^2)$ and $z = (\Pi_{ZZ}(m_Z^2) - m_Z^2 \Pi^\prime_{ZZ}(m_Z^2) )/m_Z^2$ 
\footnote{One can also extract $C$ and $z$ directly from the $Z$ mass and neutral current interaction terms in the Lagrangian. In particular, $C = 2 \left(1- {\cal O}^G_{11}\right)$ and $z = (m_{Z_1}^2 - m_Z^2)/{m_Z}^2 + C$ where the rotation matrix ${\cal O}^G$ and physical $Z$ boson mass $m_{Z_1}$ are obtained by diagonalizing the mass matrix in Eq.~(\ref{MsqZs}).}. 
All the other effective parameters $A$, $B$, $G$ and $w$ in~\cite{Burgess:1993vc} are zeros at tree level in G2HDM. 
Using Eq.~(2)  in~\cite{Burgess:1993vc} , we then obtain the tree level oblique parameters as
\bea
\label{SUEFT}
S_{\rm tree} (\{ W^\prime , X \}) & \approx & - U_{\rm tree} (\{ W^\prime , X \}) \nonumber \\ 
& \approx &   - \frac{\hat s_W^2 \hat c_W^2 }{ \hat \alpha } v^2 m_Z^2 \left[ \frac{g_H^2}{ (m_Z^2 - m^2_{W^\prime} )^2}  + \frac{g_X^2}{ ( m_Z^2 - m^2_X )^2 }  \right]  \; , \\ 
\label{TEFT}
T_{\rm tree} (\{ W^\prime , X \}) & \approx &-  \frac{1}{4 \,  \hat \alpha } v^2 \left[   
g_H^2 \frac{ (2  m_Z^2 - m^2_{W^\prime } ) }{(m_Z^2 - m^2_{W^\prime} )^2 }  
+ g_X^2 \frac{ ( 2 m_Z^2 - m_X^2 ) }{ ( m_Z^2 - m^2_X )^2 }  \right] \; .
\eea
The corresponding $W$ boson mass shift at the tree level can be computed using (\ref{wmassshift}).

\begin{figure}
	\includegraphics[width=0.6\textwidth]{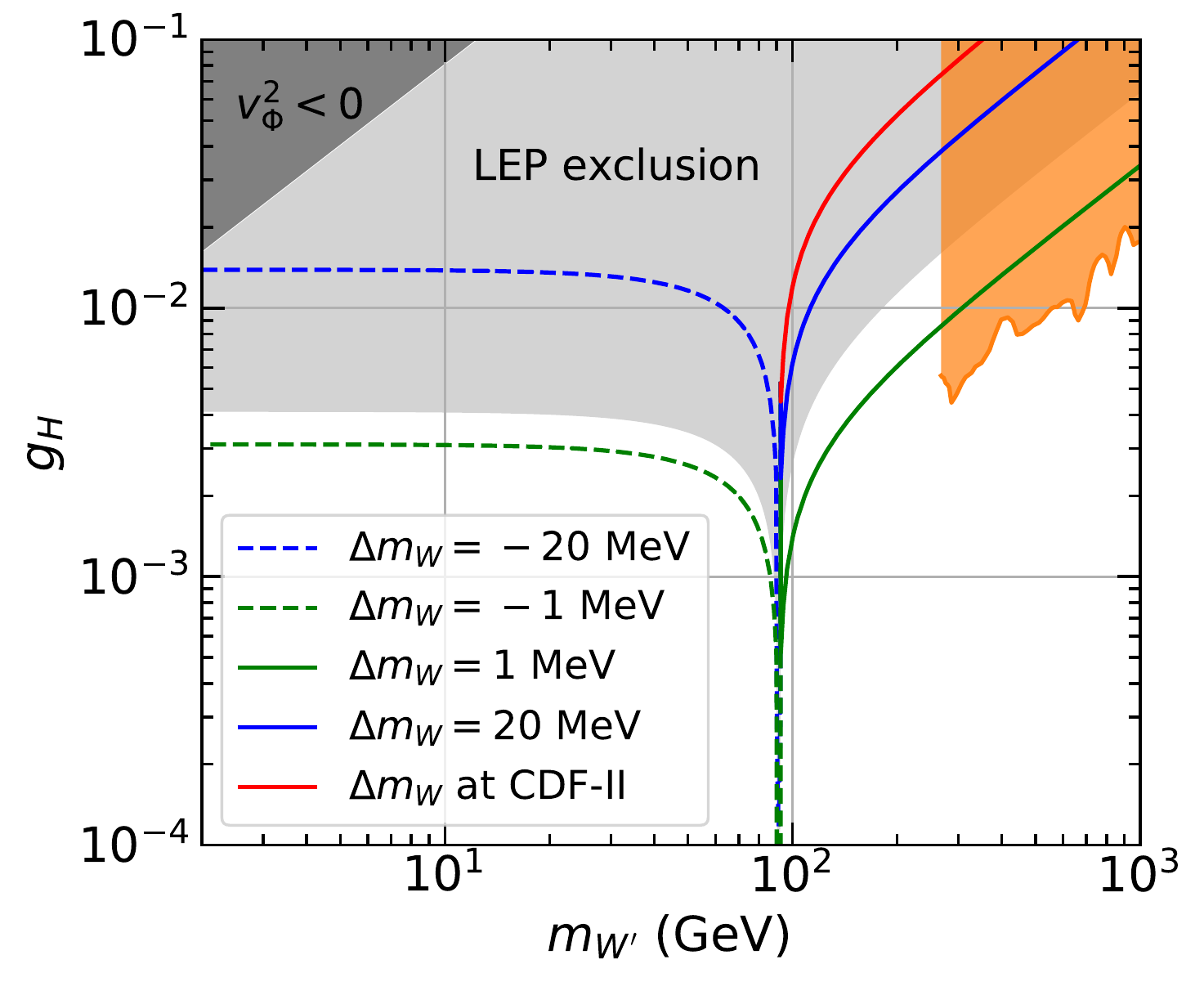}
	\caption{\label{fig:gauge} The $W$ mass shift from the tree level mixing of the neutral gauge bosons spanned on the 
	($m_{W'}$, $g_H$) plane. Here we set $g_X = 0$. The dashed green, dashed blue, solid green, solid blue and solid red lines represent the $W$ mass shift of $-1$ MeV, $-20$ MeV, $1$ MeV, $20$ MeV and the one measured at the CDF II, respectively. The light gray shaded region is the excluded region ($\Delta m_Z > 2.1$ MeV) from the measurement of $Z$ boson mass at  LEP~~\cite{Zyla:2020zbs}. The orange region is the excluded region from the di-lepton high mass resonance search from ATLAS \cite{ATLAS:2019erb}. The dark gray region is the unphysical region where $v_{\Phi}^2 < 0$, as determined by~(\ref{mwprime}).}
\end{figure}

In Fig. \ref{fig:gauge}, we show the contours of $W$ boson mass shift from the contribution of the tree level mixings between the neutral gauge bosons projected on the plane of $m_{W'}$ and $g_H$. Here we have fixed $g_X = 0$ for simplicity. Including contributions from both $W^\prime$ and $X$ do not change our conclusions in a significant way. 
For $m_{W'} \gtrsim m_Z$, the $W$ boson mass shift becomes positive values while for a lighter $m_{W'}$ the mass shift becomes negative values. 
In $m_{W'} \gtrsim m_Z$ regions and for relative large values of $g_H$, one can reach the $W$ boson mass shift ($\Delta m_W \approx 75$ MeV) measured at the CDF II. However these large values of $g_H$ also yield a large $Z$ boson mass shift which is in conflict with the great precision measurement of the $Z$ mass at LEP~\cite{Zyla:2020zbs}
\be
m_Z({\rm  LEP}) = 91.1876 \pm 0.0021 \; {\rm GeV} \; .
\ee
In the heavy mass region of ${W'}$ boson, the gauge coupling $g_H$ is also constrained to be $g_H \lesssim 10^{-2}$ from the di-lepton high mass resonance searches at ATLAS \cite{ATLAS:2019erb}.
In the allowed region, the $W$ boson mass shift is relatively small from these extra massive neutral gauge bosons in G2HDM. 
In other words, the mass mixing effects from (\ref{MsqZs}) to the $W$ boson mass shift is not significant.

In the next three subsections, we will turn to one-loop contributions to the oblique parameters from all other new particles in G2HDM. 
We will treat $Z_1 = Z  \approx Z_{\rm SM}$ and work in the Feynman-'t Hooft gauge. 

\subsection{Contributions from the $SU(2)_H$ doublet $\Phi_H$}
\label{subsec4b}

\begin{figure}[tb] 
  \includegraphics[width=13cm,height=2.6cm]{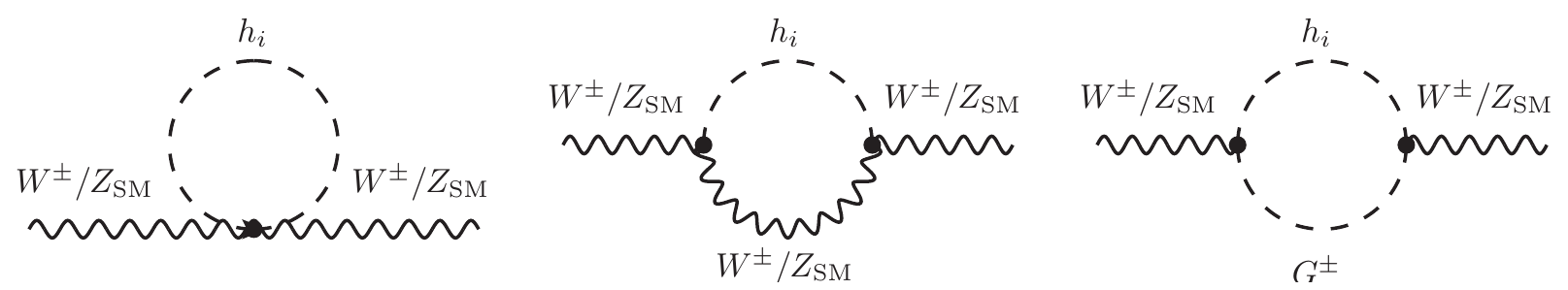}  
   \caption{Vacuum polarization diagrams that contribute to the oblique parameters from the mixing between the SM doublet $H_1$ and the hidden doublet $\Phi_H$ in G2HDM.}
   \label{STU-HiddenDoublet}
\end{figure}

Since $\Phi_H$ is a $SU(2)_H$ doublet but a $SU(2)_L$ singlet, its contribution can only arise from 
the mixing effects between $h_{\rm SM}$ and $\phi_2$ and therefore similar to the singlet extension of the SM.
The relevant Feynman diagrams are depicted in Fig.~\ref{STU-HiddenDoublet}.

For the $\Delta T$ parameter, we find
\bea
\label{eq:T-singlet}
\Delta T (\Phi_H) & = & \frac{3 \sin^2 \theta_1}{16 \pi \hat s_W^2 } 
\biggl[ 
 \frac{m_{h_2}^2}{m_{h_2}^2 - m_W^2} \log \left( \frac{m_{h_2}^2}{m_{W}^2} \right) - \left(  \frac{m_Z^2}{m_W^2} \right) \frac{m_{h_2}^2}{m_{h_2}^2 - m_Z^2}  
\log \left( \frac{m_{h_2}^2}{m_{Z}^2} \right) \biggr. \nonumber \\
&& \quad \quad \quad \quad  \biggl.  - \left( h_2 \to h_1 \right) \biggr] \; .
\eea
The above result agrees with~\cite{Profumo:2007wc,Lopez-Val:2014jva}.

Compact expressions for the $\Delta S (\Phi_H) $ and $\Delta U (\Phi_H)$ parameters can also 
be obtained using their definitions given in (\ref{S}) and (\ref{U}).
\bea
\label{eq:S-singlet}
\Delta S (\Phi_H) & = & \frac{\sin^2 \theta_1}{12 \pi} \left\{
- \frac{ 2  \, m_Z^2 \left( m_{h_1}^2 - m_{h_2}^2 \right) \left( 2 m_{h_1}^2 m_{h_2}^2  + 3 m_Z^2 \left( m_{h_1}^2 + m_{h_2}^2 \right) - 8 m_Z^4 \right) }
 { \left( m_{h_1}^2 - m_Z^2 \right)^2 \left( m_{h_2}^2 - m_Z^2 \right)^2 } \right. \nonumber \\
&+& \left. \left[ \frac{ m_{h_2}^2  \left( m_{h_2}^4 - 3 m_{h_2}^2 m_Z^2 + 12 m_Z^4 \right)  }{ \left( m_{h_2}^2 - m_Z^2 \right)^3 } \log \left( \frac{m_{h_2}^2 }{ m_Z^2 }  \right)   
- \left( m_{h_2} \to m_{h_1} \right)  \right]  \right\} \, ,
\eea
and
\bea
\label{eq:U-singlet}
\Delta U (\Phi_H) & = & \frac{\sin^2 \theta_1}{12 \pi} \Biggl\{ \Biggl[
  \frac{ 2 \,  m_Z^2 \left( m_{h_1}^2 - m_{h_2}^2 \right) \left( 2 m_{h_1}^2 m_{h_2}^2 
+ 3 m_Z^2 \left( m_{h_1}^2 + m_{h_2}^2 \right)   - 8 m_Z^4 \right) }
 { \left( m_{h_1}^2 - m_Z^2 \right)^2 \left( m_{h_2}^2 - m_Z^2 \right)^2 }  \Biggr. \Biggr. \nonumber \\
& - &  \Biggl.  \frac{ m_{Z}^4 \left( 9 m_{h_2}^2  +  m_Z^2 \right) }{ \left( m_{h_2}^2 - m_Z^2 \right)^3 } 
\log \left( \frac{m_{h_2}^2 }{ m_Z^2 }  \right)  
+ \frac{ m_{Z}^4 \left( 9 m_{h_1}^2  +  m_Z^2 \right) }{ \left( m_{h_1}^2 - m_Z^2 \right)^3 } 
\log \left( \frac{m_{h_1}^2 }{ m_Z^2 }  \right)    \Biggr] \nonumber \\
& & \quad \quad \quad -  \biggl. \biggl[ \left( m_{Z} \to m_{W} \right) \biggr]  \biggr\} \; .  
\eea
As one expects, all three $\Delta S(\Phi_H)$, $\Delta T(\Phi_H)$ and $\Delta U(\Phi_H)$ vanish as $m_{h_2} \rightarrow m_{h_1}$.

\begin{figure}
	\includegraphics[width=0.49\textwidth]{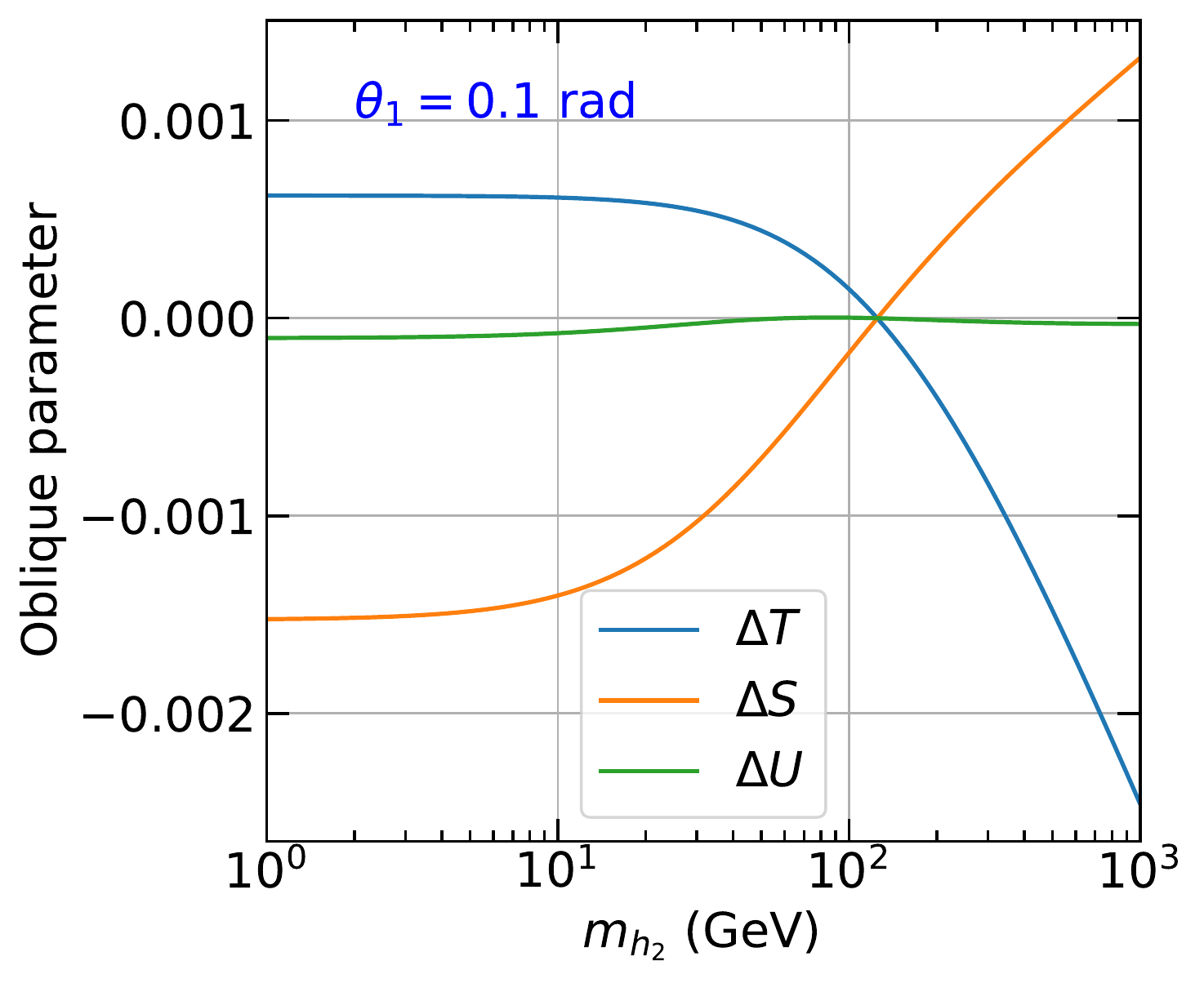}
	\includegraphics[width=0.49\textwidth]{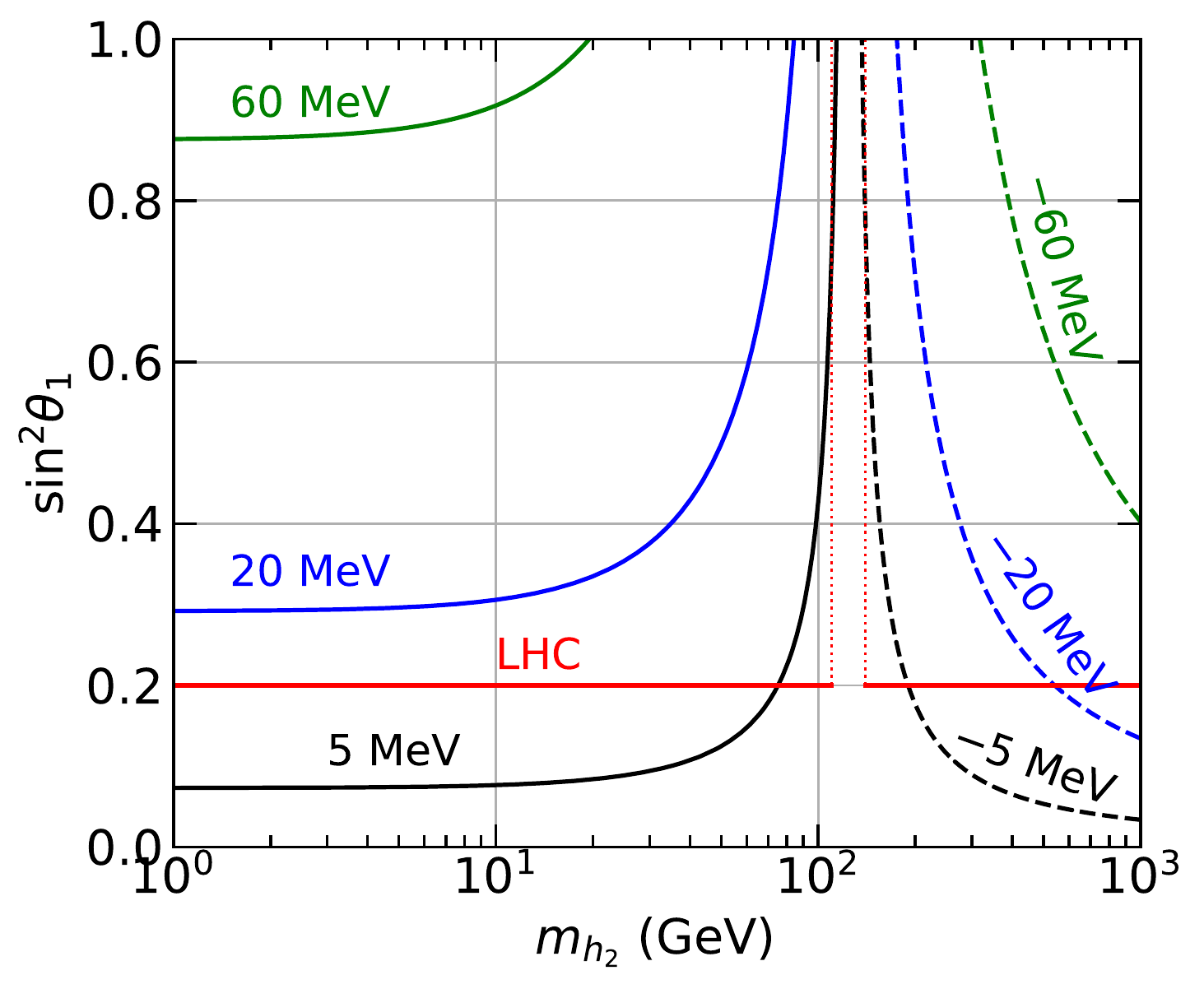}
	\caption{\label{fig:singlet} Left panel: the oblique parameters as a function of $m_{h_2}$ with fixed $\theta_1 = 0.1$ (rad). Right panel: contours for the $W$ boson mass shift, $\Delta m_W$, projected on the ($m_{h_2}$, $\sin^2\theta_1$) plane. The solid black, blue, and green 
	(dashed black, blue and green) lines represent the mass shift $\Delta m_W =$ 5, 20 and 60 ($-$5, $-$20 and  $-$60) MeV, respectively. The solid red line 
	indicates the upper bound on the mixing angle $\sin^2\theta_1 < 0.2$ from the Higgs signal strength measurement at the LHC. We note this upper bound for $\sin^2 \theta_1$ changes if $m_{h_2}$ is close to $m_{h_1}$ \cite{Robens:2015gla}.}
\end{figure}

In the left panel of Fig.~\ref{fig:singlet}, we show the oblique parameters calculated from~(\ref{eq:T-singlet}), (\ref{eq:S-singlet}), 
and (\ref{eq:U-singlet}) as a function of $m_{h_2}$ with the mixing angle $\theta_1$ fixed to be  0.1 rad. 
The parameters $\Delta S$ and $\Delta T$ flip their signs when $m_{h_2}$ passing the value of $m_{h_1}=125.38$ GeV. 
For $m_{h_2} > m_{h_1}$,  $\Delta S$ is positive value while $\Delta T$ is opposite.  One can see that $|\Delta U|$ is much smaller 
than $|\Delta S|$ and $|\Delta T|$.  

In the right panel of Fig.~\ref{fig:singlet}, we show the contours for the $W$ mass shift projected on the ($m_{h_2}$, $\sin^2\theta_1$) plane. 
For $m_{h_2} < m_{h_1}$, the $SU(2)_H$ doublet $\Phi_H$ contribution gives a positive $W$ boson mass shift, while for $m_{h_2} > m_{h_1}$, it gives a negative mass shift.
Due to the Higgs data at the LHC which require $\sin^2 \theta_1 \lesssim 0.2$ \cite{ATLAS:2021vrm} (from the Higgs boson coupling modifier $\kappa_Z = 0.99\pm0.06$), the $W$ mass shift from this contribution is constrained to be relatively small and thus makes it difficult for the hidden $SU(2)_H$ doublet $\Phi_H$ to explain the CDF $W$ mass anomaly.

\begin{figure}[tb] 
  \includegraphics[width=13cm,height=5.4cm]{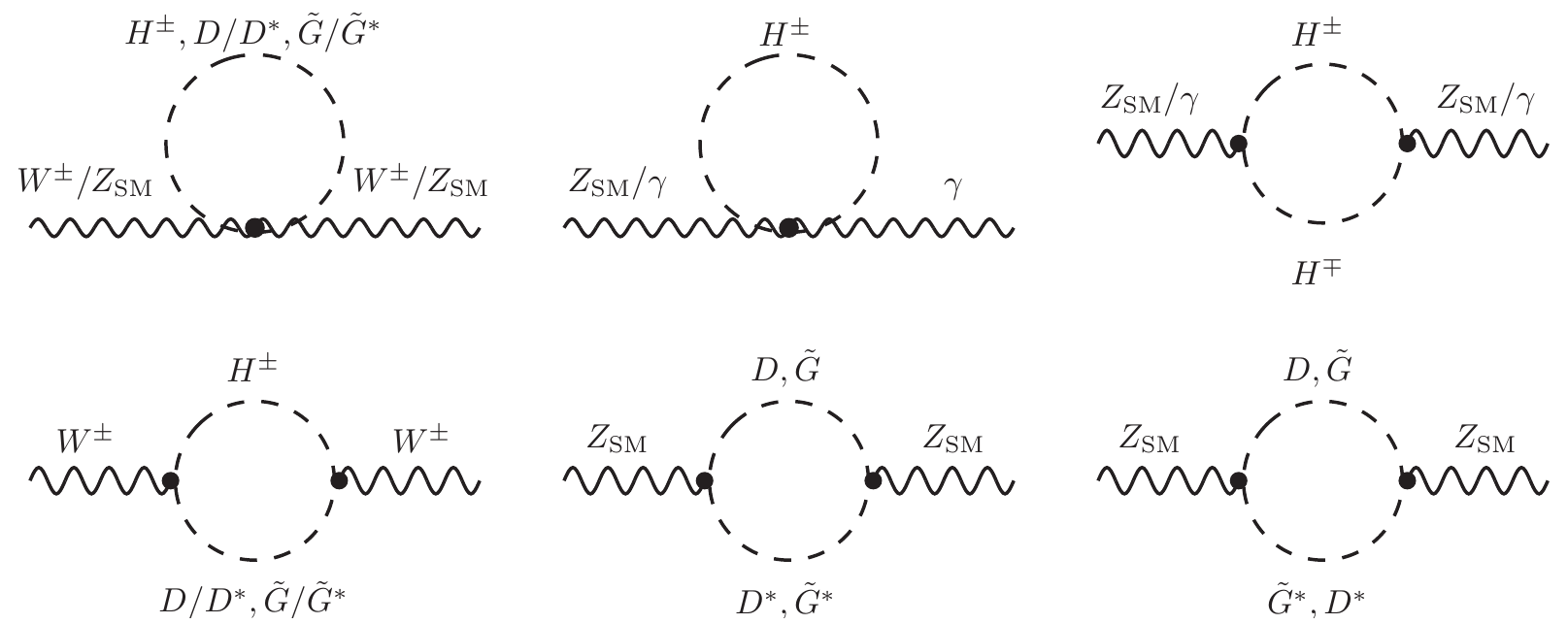} 
   \caption{Vacuum polarization diagrams that contribute to the oblique parameters from the inert doublet $H_2$ in G2HDM.}
    \label{STU-InertDoublet}
\end{figure}

\subsection{Contributions from the inert Higgs doublet $H_2$ in G2HDM}
\label{subsec4c}

The second Higgs doublet $H_2$ in G2HDM plays a similar role as the inert Higgs doublet in I2HDM~\cite{Deshpande:1977rw,Barbieri:2006dq,Arhrib:2013ela,Kephart:2015oaa}. 
However the neutral component $H_2^0$ of $H_2$ can be treated as a complex scalar field in G2HDM instead of decomposing into 
$(S+iA)/\sqrt 2$ in I2HDM, where $S$ and $A$ are the scalar and pseudo-scalar fields. In addition, as already given in (\ref{H20field}), 
$H_2^0 (H_2^{0 \, *})$ mixes with $G^m_H (G^p_H)$ from the hidden doublet $\Phi_H$ to form a physical dark Higgs $D (D^*)$ and 
a Goldstone boson $\tilde G (\tilde G^*)$, with the latter of which absorbed by the complex gauge fields $W^{\prime \,m} (W^{\prime \,p})$ of $SU(2)_H$.
The relevant Feynman diagrams for the inert Higgs doublet contributions are depicted in Fig.~\ref{STU-InertDoublet}.

For the $T$ parameter, we got
\bea
\Delta T ( H_2) & = & \frac{1 }{8 \pi^2 \hat \alpha v^2 } 
 \biggl[ F(m_{H^{\pm}},m_D)\cos^2 \theta_2 + F(m_{H^{\pm}},m_{W'})\sin^2 \theta_2  \biggr. \nonumber \\
&& \quad \quad \quad \quad  \biggl. - \,  \frac{1}{4} \, F(m_D,m_{W'})\, \sin^2 2\theta_2 \biggr]  
 \; .
\eea
where $m_{W'}$ is the mass the Goldstone boson $\tilde{G}$ which is absorbed by the new gauge boson $W^{\prime \, (p,m)}$ of $SU(2)_H$,
and the function $F(m_1,m_2)$ is defined as 
\be
F(m_1,m_2) = 
\begin{cases}
\frac{m_1^2 + m_2^2}{2} - \frac{m_1^2 m_2^2}{m_1^2 - m_2^2} \log\left(  \frac{m_1^2}{m_2^2} \right) \;\;\;\;\;\;\, \mathrm{for} \; m_1 \neq m_2   \; , \\ 
0  \;\;\;\;\; \;\;\;\;\ \;\;\;\;\ \;\;\;\;\;  \;\;\;\;\ \;\;\;\;\ \;\;\;\;\; \;\;\;\; \;\;\;\; \;   \mathrm{for} \; m_1 = m_2  \; .
\end{cases}
\ee

Again, analytical formulas for the $\Delta S (H_2) $ and $\Delta U (H_2) $ parameters can be obtained
using their definitions given in (\ref{S}) and (\ref{U}).
\bea
\Delta S (H_2) & = & \frac{1}{36 \pi} \left\{ 
 - 3 \log \left( \frac{ m_{H^\pm}^2 }{ m_D^2 } \right) + 6 \left(\cos^4\theta_2 - 1 \right) \right. \nonumber \\
&+& \biggl. 3  \left( 2 - \log \left( \frac{ m_{D}^2 }{ m_{W'}^2 } \right) \right) \sin^4\theta_2 +  \frac{1}{4} G ( m_D, m_{W'} ) \sin^2 2 \theta_2 \biggr\} 
\eea
and
\bea
\Delta U (H_2) & = & \frac{1}{36 \pi} \left\{
- 3 \log \left( \frac{ m_{H^\pm}^2 }{ m_D^2 } \right) 
- 6 \left(\cos^4\theta_2 + 1 \right) +  G \left(  m_D , m_{H^\pm}  \right) \cos^2\theta_2  
\right. \nonumber \\
& - & 3 \left( 2 - \log \left( \frac{m_D^2}{m_{W'}^2} \right) \right) \sin^4 \theta_2 \nonumber \\
& - & \left( \, 6 \log \left( \frac{m_D^2}{m_{W'}^2} \right) - G ( m_{W'} , m_{H^\pm} ) \right) \sin^2\theta_2 \nonumber \\
& - & \left. \frac{1}{4} G \left( m_D, m_{W'} \right)  \sin^2 2 \theta_2 \right\} \; ,
\eea
where
\bea
G \left( m_1, m_2 \right) & = &  \frac{\left( 7 m_1^4 - 2 m_1^2 m_2^2 + 7 m_2^4 \right) }{\left( m_1^2 - m_2^2 \right)^2}  - 
\, 6 \, \frac{ m_2^4 \left( 3 \, m_1^2 -  m_2^2  \right) } {\left( m_1^2 - m_2^2 \right)^3}  \log \left( \frac{m_1^2}{m_2^2} \right) \; .
\eea
Note that in the limit of $m_2 \to m_1$, $G (m_1,m_1)=12$.

In the pure inert limit of $\theta_2 \to 0$, $D \to {H_2^0}$ and we simply have
\bea
\label{S-Inert}
\lim_{\theta_2 \to 0} \Delta S(H_2) & = & - \frac{1}{12 \pi} \log \left( \frac{m_{H^\pm}^2}{m_{H_2^0}^2} \right) \;, \\
\label{T-Inert}
\lim_{\theta_2 \to 0} \Delta T ( H_2) & = & \frac{1 }{8 \pi^2 \hat \alpha v^2 } F(m_{H^{\pm}},m_{H_2^0}) \; , \\
\label{U-Inert}
\lim_{\theta_2 \to 0} \Delta U(H_2) & = & - \frac{1}{12 \pi} \log \left( \frac{m_{H^\pm}^2}{m_{H_2^0}^2} \right) 
 + \frac{1}{36 \pi} \left( G \left(  m_{H_2^0} , m_{H^\pm} \right)  - 12 \right)  \; .
\eea
The above expressions of (\ref{S-Inert}) and (\ref{T-Inert}) are consistent with the inert Higgs results~\cite{Barbieri:2006dq}.
Furthermore, if $m_{H^\pm}=m_{H_2^0}$, $\Delta S(H_2) = \Delta T(H_2) = \Delta U(H_2) = 0$ in this limit. 

\subsection{Contributions from the new heavy fermions in G2HDM}
\label{subsec4d}

Since all the new heavy fermions $f^H$ in G2HDM are $SU(2)_L$ singlets, they don't interact with the charge $W^\pm$ bosons. 
Under the assumption that $g_H, g_X \ll g, g^\prime$, the heavy fermions interacts with both the SM $\gamma$ and $Z$ 
are vector-like described by the following Lagrangian
\be
\mathcal L ( f^H ) = e Q_{ f^H } \left(  \bar f^H \gamma_\mu f^H  \right) \left( A^\mu -  \tan \theta_W Z^\mu \right) + \cdots \; ,
\ee
where we have dropped terms that are proportional to $g_H$ or $g_X$.
Thus we have
\bea
\label{PiWW}
\Pi_{WW}^{ f^H} ( q^2 ) & = & 0 \; ,  \\
\label{Pigg}
\Pi_{ \gamma \gamma }^{ f^H} ( q^2 ) & = & N_C e^2 Q^2_{f^H} \Pi_{QQ} ( q^2 ) \; , \\ 
\label{PigZ}
\Pi_{ \gamma Z }^{ f^H} ( q^2 ) & = & - N_C e^2 Q^2_{f^H} \tan \theta_W  \Pi_{QQ} ( q^2 ) \; , \\
\label{PiZZ}
\Pi_{ ZZ }^{ f^H} ( q^2 ) & = & N_C e^2 Q^2_{f^H} \tan^2 \theta_W \Pi_{QQ} ( q^2 ) \; , 
\eea
where $\Pi_{QQ}$ is the oblique loop amplitude $\Pi_{\gamma\gamma}$ with both the color and electric charge factors trimming off~\cite{Peskin:1995ev}, {\it i.e.}
\be
\label{PiQQ}
\Pi_{QQ} ( q^2 )  = \frac{1}{2 \pi^2} q^2 \left( \frac{1}{6} E - \int_0^1 dx \, x (1-x) \log \frac{m^2_{f^H} - x \left( 1 - x \right) q^2}{ \mu^2} \right)
\ee
with $E \equiv \frac{2}{\epsilon} - \gamma_E + \log \left( 4 \pi / \mu^2 \right)$. 
Using the above expressions (\ref{PiWW})-(\ref{PiQQ}), 
we can demonstrate easily that all the oblique parameters from the heavy fermions $f^H$s in G2HDM vanish:
\be
\label{STUfH}
\Delta S (f^H) = \Delta T (f^H) = \Delta U (f^H) = 0 \; .  
\ee
The non-trivial leading contributions from the heavy fermions in G2HDM to the oblique parameters start at order 
$g_H^2/16 \pi^2$ and $g_X^2/16\pi^2$, which we are neglecting in this study.
 
To summarize, we have computed all possible nontrivial sources of new physics effects 
to the oblique parameters $\Delta S$, $\Delta T$ and $\Delta U$ in G2HDM under the approximation of 
$g_H, g_X \ll g, g^\prime$. While the tree level mixings of the neutral gauge bosons and the extra heavy fermions in G2HDM 
contribute to the oblique parameters are of order $g_{H,X}^2$ and $g_{H,X}^2/16\pi^2$ respectively and therefore not significant, 
there are new contributions from the inert doublet $H_2$ and the $H_2$-$\Phi_H$ mixings,
as well as from the hidden doublet $\Phi_H$ through the $H_1$-$\Phi_H$ mixings. These new contributions from the extended scalar sector in G2HDM
are of order $g^2/16\pi^2$ and $g^{\prime 2}/16\pi^2$ which are the same order as the SM  one-loop  contributions. In the next section, we will focus on 
the detailed numerical analysis of these new contributions.

\begin{table}[htbp]
\begin{tabular}{|c|cc|}
\hline
\multirow{2}{*}{Parameter  [units]} & \multicolumn{2}{c|}{Range/value [scan prior]}    \\ \cline{2-3} 
                  & \multicolumn{1}{c|}{Light DM mass scenario} & Heavy DM mass scenario \\ \hline
                  $m_{W^{\prime}}  $ [GeV]  & \multicolumn{1}{c|}{$[0.01\,,\,50]$ [log]} & $ [100\,,\,2000]$ [linear] \\
                  $M_{X}$ [GeV] & \multicolumn{1}{c|}{$[0.01\,,\,100]$ [log]} & $ 3000 $   \\
                  $g_{X}$ & \multicolumn{1}{c|}{ $[10^{-6}\,,\,0.1]$ [log] } &  $ 10^{-5}$ \\ \hline
                  $m_{h_2}$ [GeV]  & \multicolumn{2}{c|}{$[m_{h_1}\,,\,2000]$  [linear] }    \\
                  $m_{H^{\pm}}$ [GeV]  & \multicolumn{2}{c|}{$ [100\,,\,2000]$ [linear]}    \\
                  ($m_{H^{\pm}} - m_D$) [GeV]  & \multicolumn{2}{c|}{ $[-500, 500]$ [linear] }    \\
                  $\theta_{1,2}$ [rad] & \multicolumn{2}{c|}{$[-\frac{\pi}{2} , \frac{\pi}{2}]$   [linear]}    \\ \hline
\end{tabular}
\caption{\label{tab:prior}
The parameter space setup for the scan. For the light DM mass scenario, $m_{W^{\prime}}$, $m_{X}$ and $g_{X}$ are scanned in log scale while the rest are in linear scale. 
All the new heavy fermion masses are set equal to 3 TeV. 
}
\end{table}

\section{Numerical Results}
\label{sec5}

In this section we present the numerical results in the light of new $W$ boson mass measurement at the CDF II. 
Among the $h$-parity odd particles, we require $W'$ to be the lightest particle 
so that it can be a DM candidate in this model. 
We propose two setups of scan based on the mass of the DM, one is
the light DM mass scenario and another is the heavy DM mass scenario. 
The parameter space setup for these two scenarios is given in Table~\ref{tab:prior}. 
We sample the parameter space in the model using MCMC scans {\tt emcee} 
\cite{ForemanMackey:2012ig}. 
For the light DM mass scenario, 
$m_{W^{\prime}}$, $M_{X}$ and $g_{X}$ are scanned in the log scale, 
while the rest  are in the linear scale in both scenarios except $M_{X}$ and $g_X$ are fixed in the heavy DM mass scenario.
We also assume that all heavy fermion masses are degenerated and fixed to be $3$ TeV.

We closely follow the analysis for the current constraints in the model from~\cite{Ramos:2021omo, Ramos:2021txu}. 
In particular, we take into account the theoretical constraints on the scalar potential, 
the collider physics from the LHC including the signal strengths 
of $h \to \gamma \gamma$~\cite{Aad:2019mbh} and  $h \to \tau^+ \tau^-$~\cite{Sirunyan:2018koj} from the gluon-gluon fusion, the constraints from the electroweak precision measurement at $Z$ pole~\cite{Zyla:2020zbs}
and from $Z'$~\cite{ATLAS:2019erb} and dark photon $\gamma'$ physics (see~\cite{Fabbrichesi:2020wbt} for a recent review).  
To take into account the new $W$ boson mass measurement at the CDF II, 
we adopt the recent global fit values for the oblique parameters from~\cite{deBlas:2022hdk},
which are given as
\bea
\label{eq:STUnew}
S &=& 0.005 \pm 0.096 \; , \nonumber \\
T &=& 0.04 \pm 0.12 \; ,  \\
U &=& 0.134 \pm 0.087 \; \nonumber
\eea 
and the correlation coefficients are $0.91, -0.65$ and $-0.88$ for ($S, T$), ($S, U$) and ($T,U$ ), respectively. 

We also take into account the constraints from DM searches including the DM relic density 
 $\Omega {\rm h}^2 = 0.120 \pm 0.001$ measured from Planck collaboration \cite{Aghanim:2018eyx}, 
DM direct detections from CRESST III \cite{Angloher:2017sxg}, DarkSide-50 \cite{Agnes:2018ves}, XENON1T \cite{XENON:2018voc, Aprile:2019xxb}, PandaX-4T \cite{PandaX-4T:2021bab} and LZ \cite{LUX-ZEPLIN:2022qhg}, and the Higgs invisible decays constraint from the LHC \cite{ATLAS:2022yvh}. 
We note that, due to the kinematical forbiddance, the Higgs invisible decays constraint is not applied for the heavy DM mass scenario. The branching ratio of invisible Higgs decay is given in \cite{Ramos:2021omo, Ramos:2021txu}. 
We use micrOMEGAs package \cite{Belanger:2018ccd}
to calculate the DM relic density and the DM-proton scattering cross section. 
We note that the production cross section of the mono-jet signals $pp \to W'^{p} W'^{m} j$ in the model is small due to the smallness of the gauge coupling $g_H$ \cite{Ramos:2021omo, Ramos:2021txu} and hence evading the current constraint from the LHC \cite{Sirunyan:2017hci, ATLAS:2021kxv}.
Thus we do not include the mono-jet constraint in our analysis.

Hereafter, we denote the scan without the DM constraints as {\tt CDF-2022} and with the DM constraints as {\tt CDF-2022+DM}.
To see the impacts due to the CDF $W$ mass boson measurement, we employ other scans with the old global fit values for the oblique parameters taken from the 
Particle Data Group (PDG)~\cite{Zyla:2020zbs}
which are given as 
\bea
\label{eq:STU_pdg}
S &=& -0.01 \pm 0.1 \; , \nonumber \\
T &=& 0.03 \pm 0.12 \; ,  \\
U &=& 0.02 \pm 0.11 \; \nonumber
\eea 
and the correlation coefficients are $0.92, -0.8$ and $-0.93$ for ($S, T$), ($S, U$) and ($T,U$), respectively . 
We then denote the PDG scan without the DM constraints as {\tt PDG-2021} and with the DM constraints as {\tt PDG-2021+DM}.

\begin{figure}[tb]
	\includegraphics[width=0.49\textwidth]{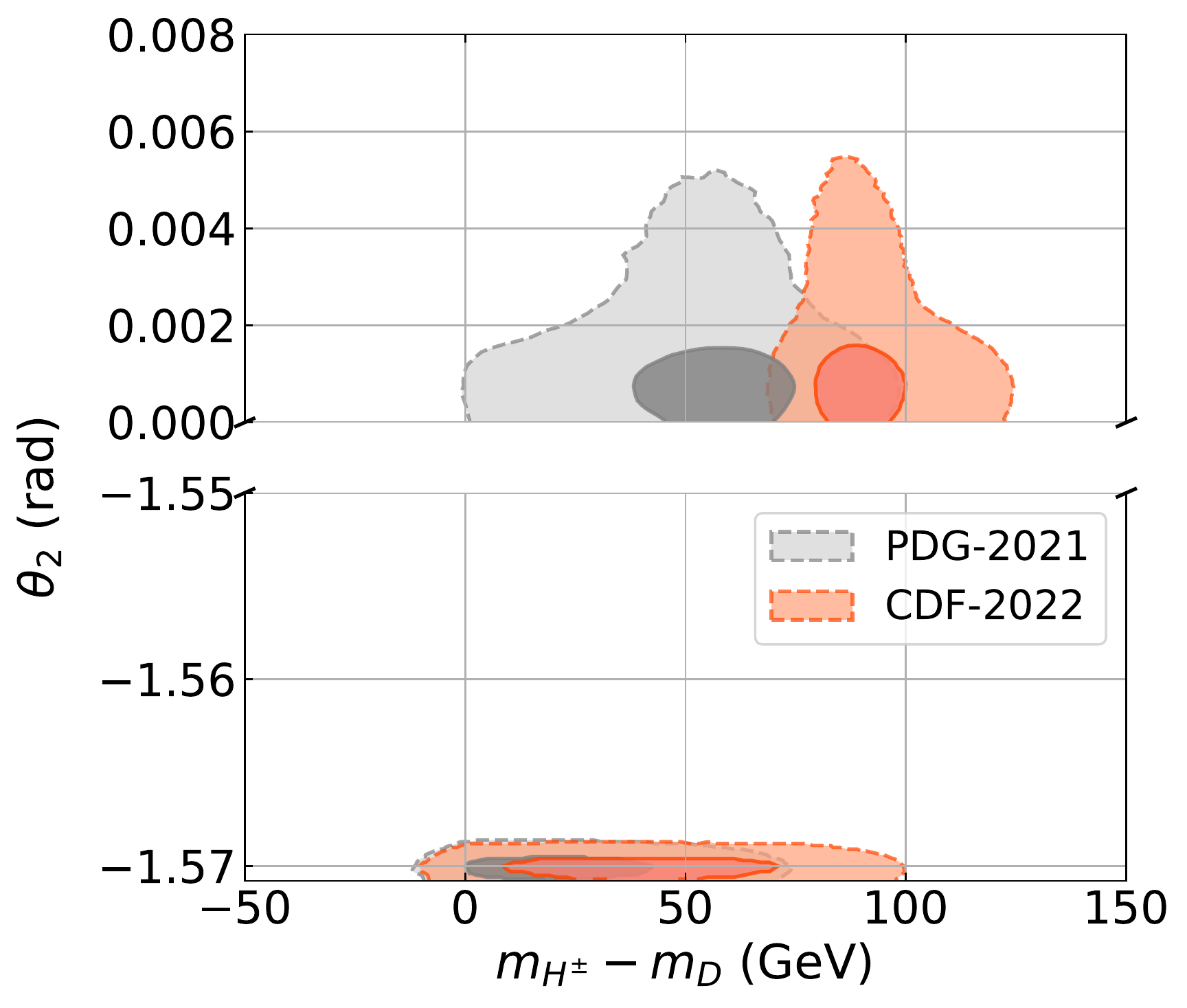}
	\includegraphics[width=0.49\textwidth]{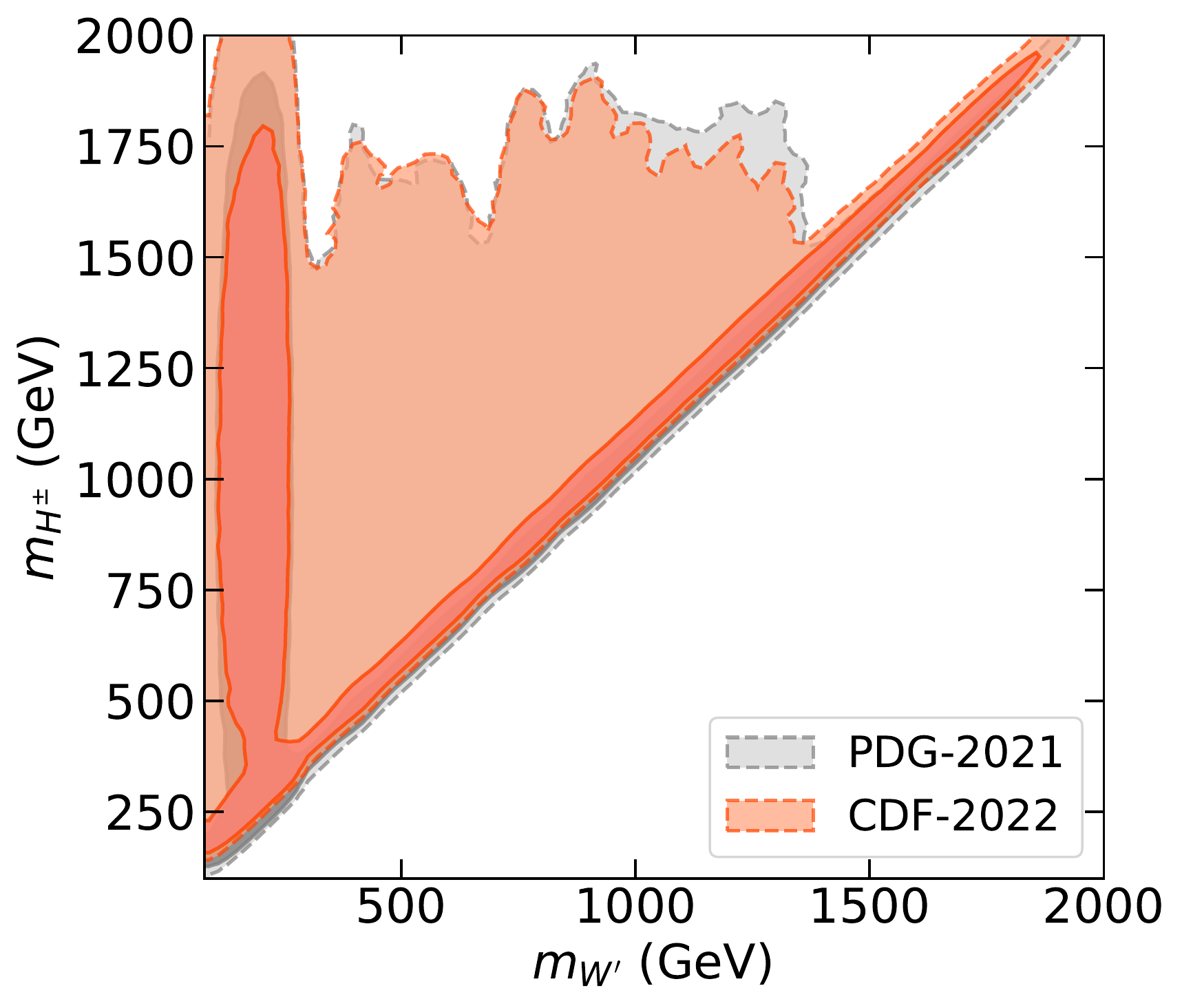}
	\caption{\label{fig:parafavored} The favored regions projected on the planes of 
	$(\Delta m \equiv m_{H^{\pm}} - m_D, \theta_2)$ (left panel) and $(m_{W^{\prime}}, m_{H^{\pm}})$ (right panel) for the heavy DM mass scenario . The dark (light) orange region represents the $1\sigma$ ($2\sigma$) favored by {\tt CDF-2022}, while the dark (light) gray region indicates the $1\sigma$ ($2\sigma$) favored by {\tt PDG-2021}.  
	}
\end{figure}

\subsection{ Heavy DM mass scenario}

In Fig.~\ref{fig:parafavored}, we show the
favored regions from {\tt CDF-2022} (orange regions) and {\tt PDG-2021} (gray regions) spanned on the parameter space. 
On the left panel of Fig.~\ref{fig:parafavored}, 
we project the favored regions on the plane of the mass splitting ($\Delta m \equiv m_{H^\pm} - m_D$) and the mixing angle $\theta_2$. 
The mixing angle $\theta_2$ is allowed to be either a nearly maximal mixing region ($\theta_2 \sim -\pi/2$) 
or a tiny mixing region ($\theta_2 \lesssim 5\times 10^{-3}$). 
Since the relation between $\theta_2$ and $g_H$ is given as 
\be
\label{eq:gHtheta2relation}
g_H  = \frac{2 \,m_{W'}}{v} \times
\begin{cases}
\begin{matrix}
 |\sin \theta_2|  \, , &&  {\rm for} \; \theta_2 > 0 \; , \\
 |\cos \theta_2| \, , &&  {\rm for} \; \theta_2 \leq 0 \; , 
\end{matrix}
\end{cases} 
\ee
the upper bound on $\theta_2$ is due to the upper bound on the gauge coupling $g_H$ which is from the constraints of the $Z$ mass shift and the di-lepton high mass resonance search at the LHC as shown in Fig. \ref{fig:gauge}. 
For the tiny mixing $\theta_2$ region, the impact from the new $W$ mass measurement at CDF II is significant. In particular, within $2\sigma$ favored region, {\tt CDF-2022} prefers a large mass splitting between the charged Higgs and dark Higgs, 
while the {\tt PDG-2021} prefers a smaller mass splitting and even allows the degenerated case. 
On the other hand, for the nearly maximal mixing region, both {\tt CDF-2022} and {\tt PDG-2021} allow the degenerated mass between the charged Higgs and dark Higgs.
However {\tt CDF-2022} still allows for a larger mass splitting as compared with the region favored by {\tt PDG-2021}.

On the right panel of Fig.~\ref{fig:parafavored}, we project the favored regions on the plane of the charged Higgs and $W'$ masses. 
The favored regions from {\tt CDF-2022} and {\tt PDG-2021} are almost the same. 
For the region of $m_{W'} \gtrsim 1.4$ TeV, the favored region has a thin cigar shape indicating 
the charged Higgs and $W'$ masses are correlated linearly and hence their mass ratio is close to unity.

\begin{figure}[tb]
	\includegraphics[width=0.6\textwidth]{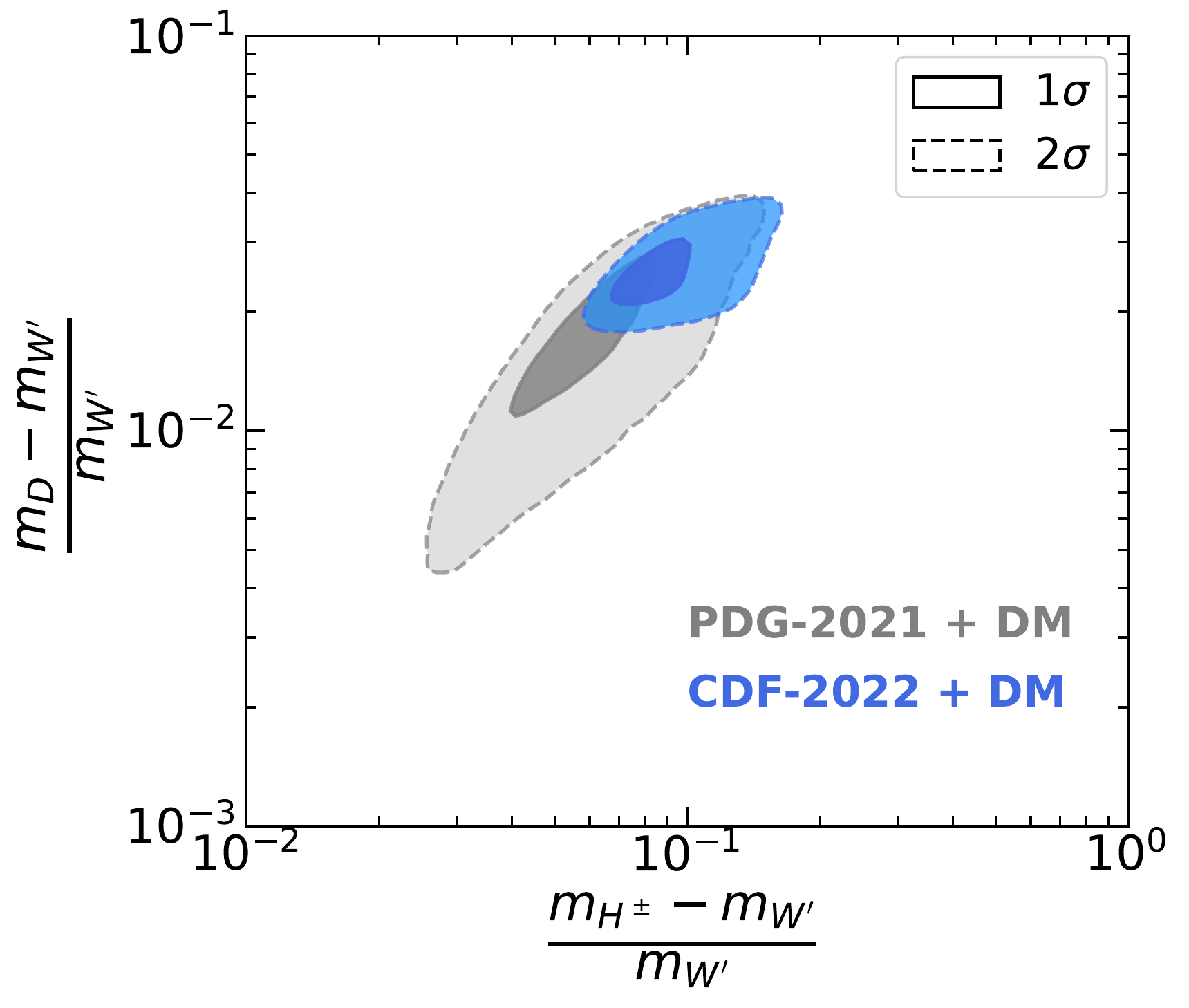}
	\caption{\label{fig:massspliting} The $1\sigma$ (dark) and $2\sigma$ (light) favored regions of the data from {\tt CDF-2022+DM}  (blue) and  {\tt PDG-2021+DM}  (gray)  projected on the plane of the mass splittings $(m_{H^\pm} - m_{W^{\prime}})/{m_{W^{\prime}}}$ and $(m_{D} - m_{W^{\prime}})/{m_{W^{\prime}}}$ for the heavy DM mass scenario. }
\end{figure}

\begin{figure}
	\includegraphics[width=0.49\textwidth]{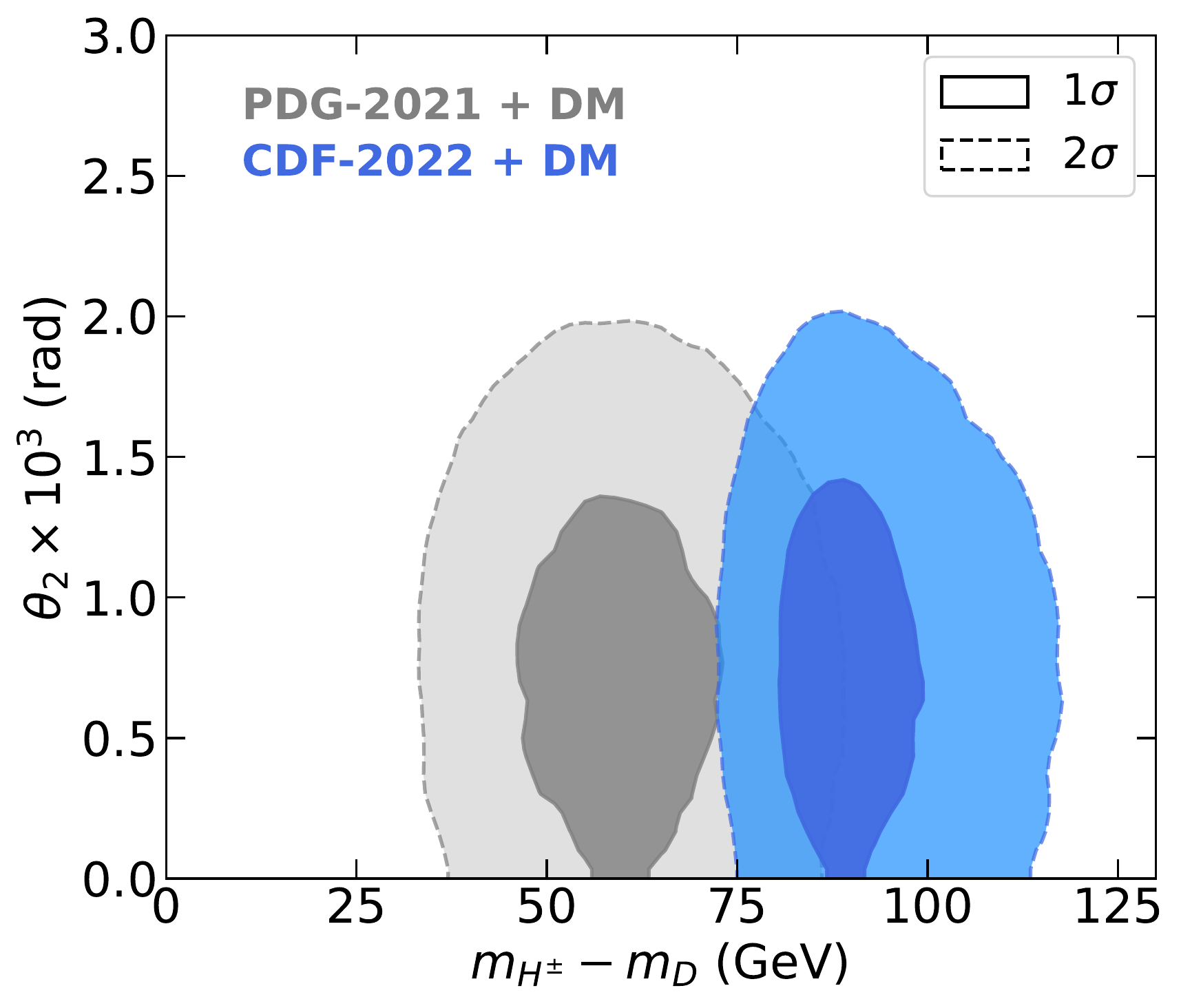}
	\includegraphics[width=0.49\textwidth]{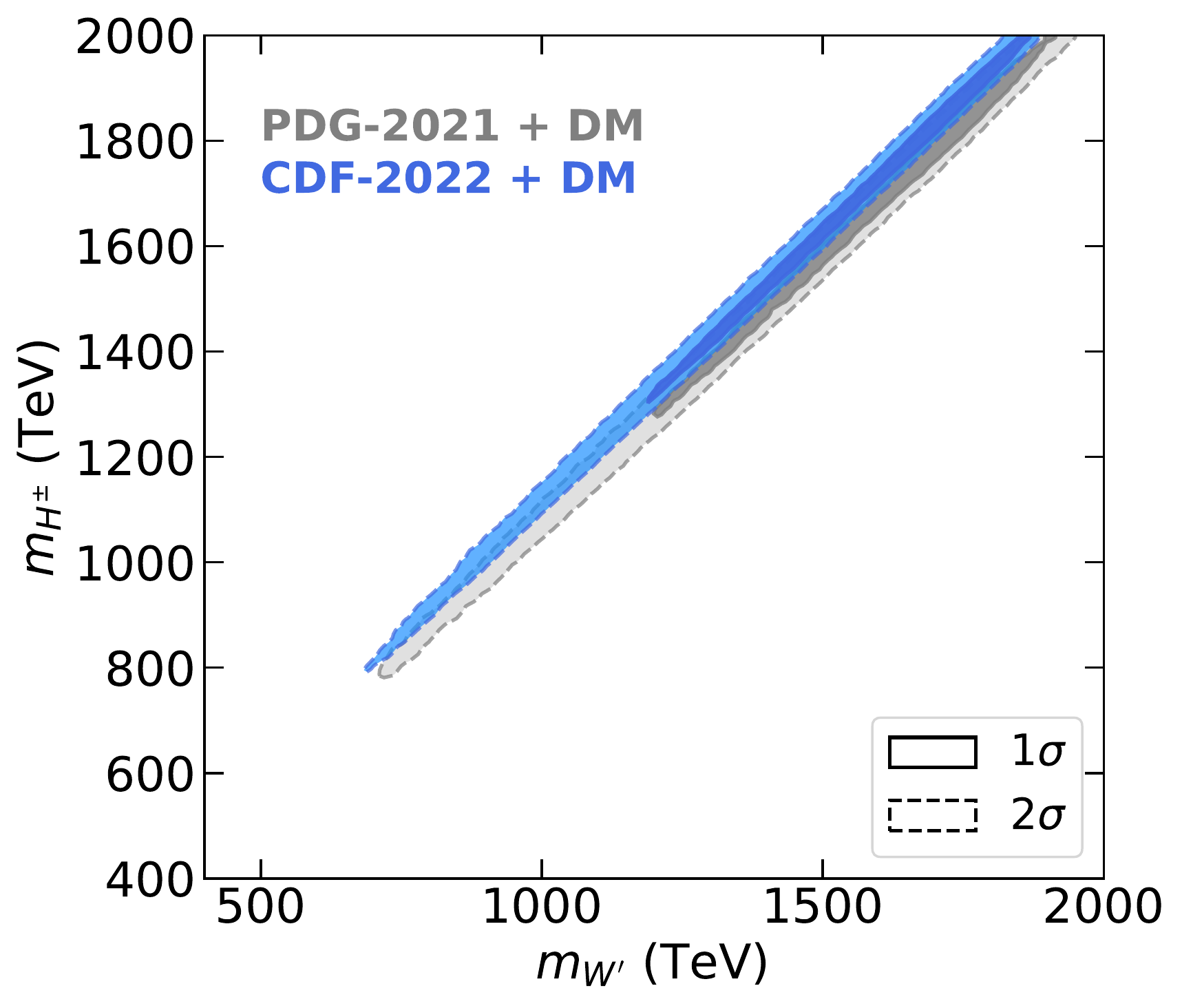}
	\caption{\label{fig:DMrelicparafavored} The $1\sigma$ (dark) and $2\sigma$ (light) favored regions of the data from {\tt CDF-2021+DM}  (blue) and  {\tt PDG-2022+DM}  (gray) projected 
	on the planes of 
	$(\Delta m \equiv m_{H^{\pm}} - m_D, \theta_2)$ (left panel) and $(m_{W^{\prime}}, m_{H^{\pm}})$ (right panel)  for the heavy DM mass scenario.  }
\end{figure}

Similar to the well known WIMP DM scenario, the DM candidate $W'$ is kept in the chemical equilibrium with the SM thermal bath via its $2\to 2$ annihilations before starting to freeze-out due to the expansion of the universe. 
The DM relic abundance can be determined by solving the Boltzmann equation for the evolution of the DM number density which 
heavily influenced by the $2\to 2$ annihilations between the DM and SM particles. 
In G2HDM, beside the standard annihilation of the pairs $W^{\prime \, (p,m)}$ to pairs of SM particles, the coannihilation - mutual annihilation of multiple $h$-parity odd species - to pairs of SM particles can be also occurred.  The coannihilation process can be significant if the masses of $W'$ and the other $h$-parity odd particles are nearly degenerated \cite{Griest:1990kh}. 
For large mass splittings between $W'$ and other $h$-parity odd particles, 
the pairs of $W^{\prime \, (p,m)}$ mainly annihilate to pairs of SM fermions and $W^{+} W^{-}$, 
which are mediated by the neutral gauge bosons $Z, Z', \gamma'$ and scalar bosons $h_1, h_2$ via the $s$-channel as well as the new heavy fermions via the $t$-channel. 

For the heavy DM mass scenario, we found out that the main contribution that yields the DM relic density observed at Planck Collaboration is the coannihilation channels. The annihilation processes are subdominant because the cross sections are suppressed due to the smallness of the new gauge couplings $g_H$ and $g_X$. 
Fig. \ref{fig:massspliting} shows the mass difference between the DM $W'$ and $h$-parity odd particles $H^{\pm}$ and $D$ when the DM constraints are included. 
The mass differences $(m_{H^\pm} - m_{W^{\prime}})/{m_{W^{\prime}}}$ and $(m_{D} - m_{W^{\prime}})/{m_{W^{\prime}}}$ are required to be ${\cal O} (10^{-2} - 10^{-1})$ within $2\sigma$ region. 
The {\tt CDF-2022+DM} prefers a larger region of the mass differences while the {\tt PDG-2021+DM} can extend to a lower region. 

Fig. \ref{fig:DMrelicparafavored} shows the favored regions on $(\Delta m , \theta_2)$ (left panel) and $(m_{W^{\prime}}, m_{H^{\pm}})$ (right panel) 
planes when the DM constraints are taken into account. 
As compared with the results without the DM constraints (shown in Fig. \ref{fig:parafavored}), the nearly maximal mixing regions of the angle $\theta_2$ are not favored anymore. The degeneracy of the charged Higgs and dark Higgs ({\it i.e.} $\Delta m = 0$) favored by the {\tt PDG-2021} is no longer favored when the DM constraint is included. Within $2\sigma$ region, it requires the mass splitting to be in the range of $33 \,{\rm GeV} \lesssim \Delta m \lesssim  87 \,{\rm GeV}$ for {\tt PDG-2021+DM} and $72 \,{\rm GeV} \lesssim \Delta m \lesssim  118 \,{\rm GeV}$ for {\tt CDF-2022+DM}. The mixing angle is also required to be smaller $\theta_2  \lesssim  2 \times 10^{-3}$.  
The cigar shape of the favored region on the right panel of Fig. \ref{fig:DMrelicparafavored} is due to  the happenstance of DM coannihilation  {\it i.e.} $(m_{H^\pm} - m_{W^{\prime}})/{m_{W^{\prime}}} \sim {\cal O} (10^{-2} - 10^{-1})$ as suggested already in Fig. \ref{fig:massspliting} for $m_{W^\prime} \gtrsim 1.4$ TeV even before imposing DM constraints. The DM mass is predicted to be $m_{W'} > 700$ GeV while the charged Higgs mass is $m_{H^\pm} > 800$ GeV within $2\sigma$ region. The {\tt CDF-2022+DM} prefers a bit higher in the charged Higgs mass region as compared with the result from {\tt PDG-2021+DM}.

\begin{figure}[tb]
	\includegraphics[width=0.6\textwidth]{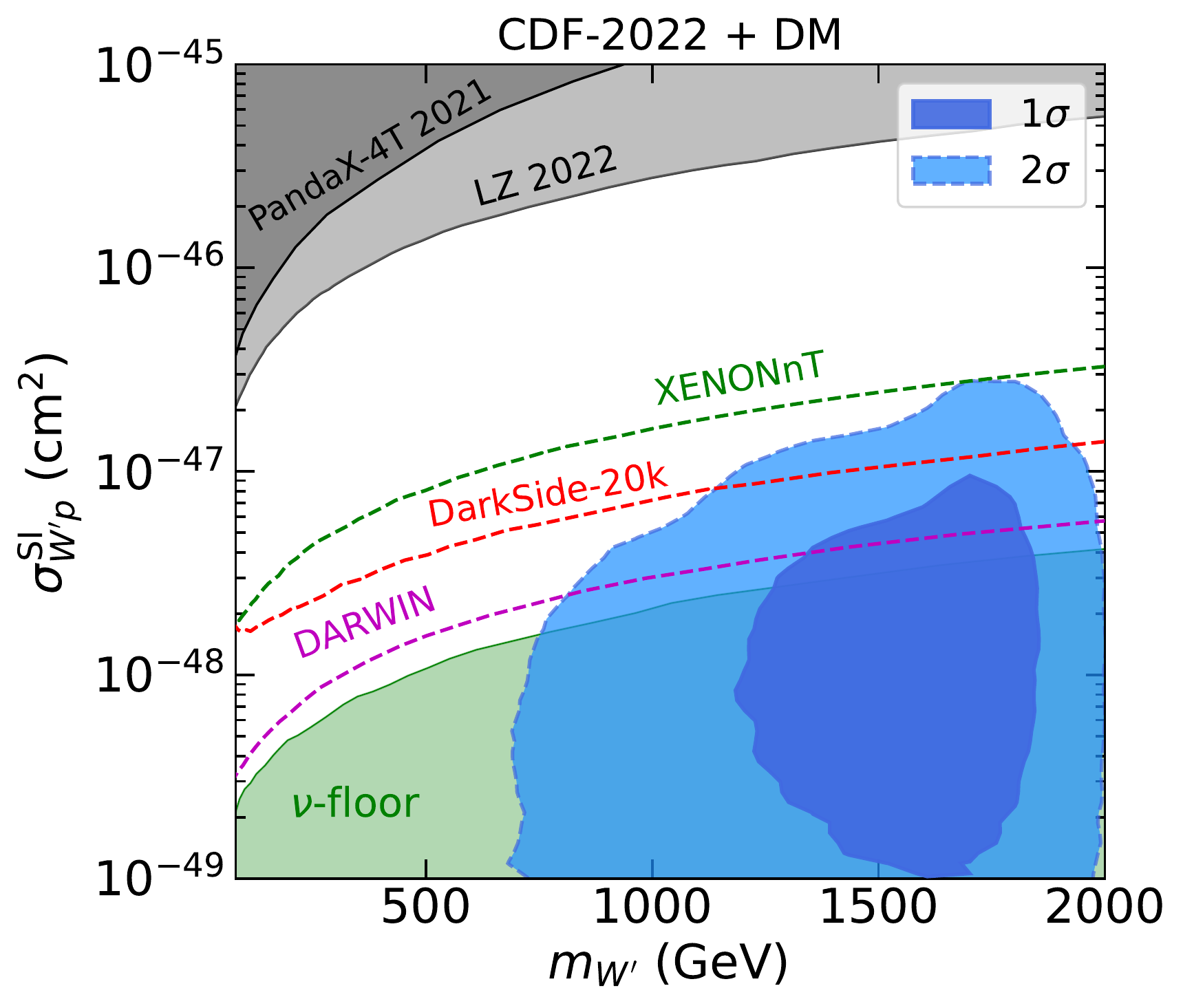}
	\caption{\label{fig:DMxsec} The $1\sigma$ (dark blue) and $2\sigma$ (light blue) favored regions of the data from {\tt CDF-2022+DM} projected on the plane of the DM mass and spin independent DM-proton scattering cross section for the heavy DM mass scenario. The dark and light gray regions are the exclusion from PandaX 4T \cite{PandaX-4T:2021bab} and LZ \cite{LUX-ZEPLIN:2022qhg} experiments . The dashed green, red and purple lines represent future sensitivities from the DM direct detections at XENONnT \cite{XENON:2020kmp}, DarkSide-20k \cite{DarkSide-20k:2017zyg} and DARWIN \cite{DARWIN:2016hyl}, respectively. Green region is the neutrino floor background.}
\end{figure}

In Fig. \ref{fig:DMxsec}, we show the {\tt CDF-2022+DM} favored region on the $(m_{W^\prime},\sigma_{W^\prime p}^{\rm SI})$ plane for the DM direct detection.
Due to the constraint from the di-lepton high mass resonance search at the LHC \cite{ATLAS:2019erb}, the gauge coupling is required to be $g_H \lesssim 10^{-2}$ in the favored DM mass region. It results in a small DM-proton spin-independent scattering cross section $\sigma_{W^\prime p}^{\rm SI}$
and thus the favored region lies far below the current limits from PandaX 4T~\cite{PandaX-4T:2021bab} and LZ~\cite{LUX-ZEPLIN:2022qhg} (gray regions). Most of the {\tt CDF-2022+DM} favored region overlaps with the neutrino floor 
background region. 
However a portion of the favored region predicted by the model can be probed by future DM direct detections at DarkSide-20k \cite{DarkSide-20k:2017zyg} and DARWIN \cite{DARWIN:2016hyl}. 

\begin{figure}[tb]
	\includegraphics[width=0.49\textwidth]{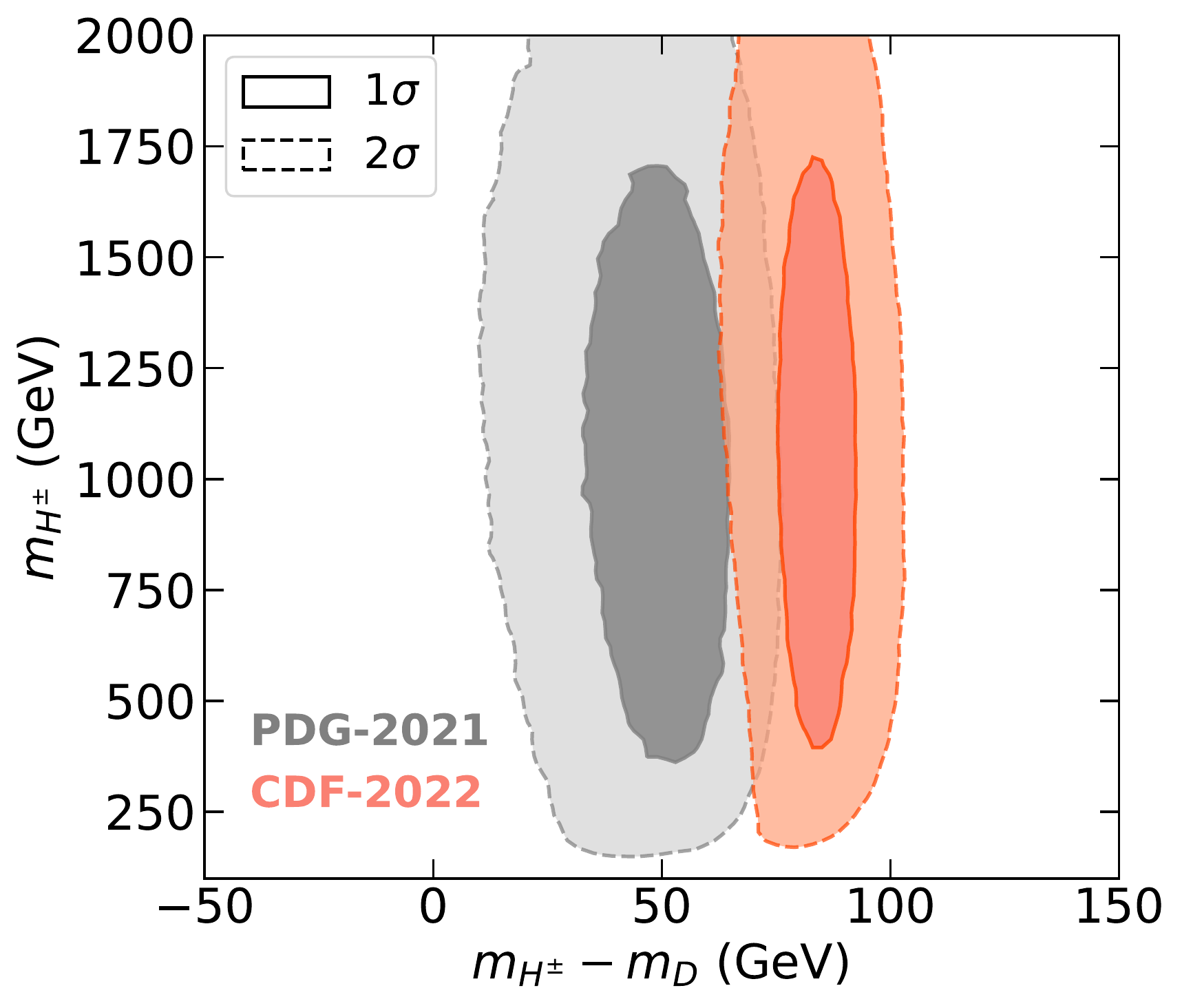}
	\includegraphics[width=0.49\textwidth]{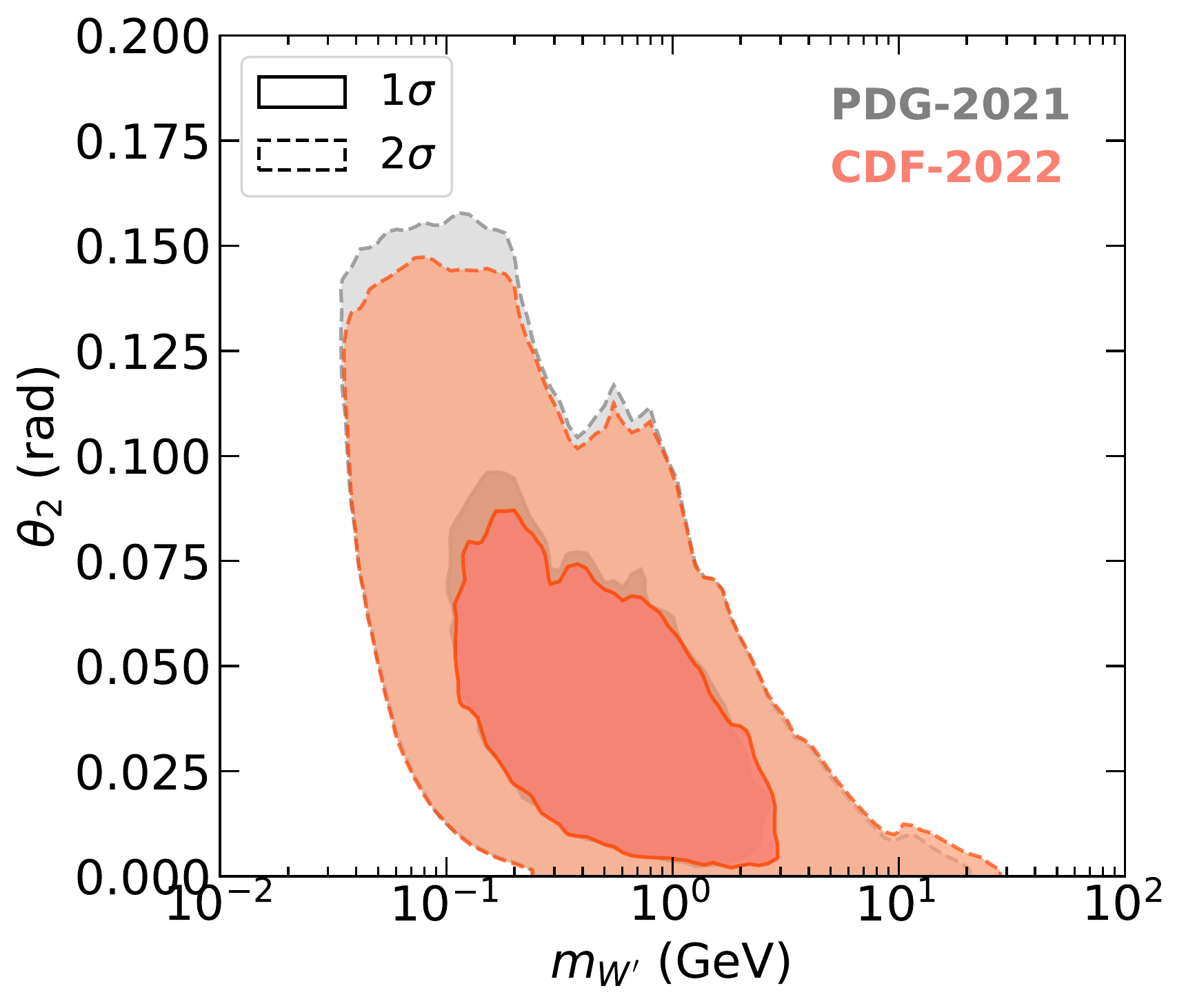}
	\caption{\label{fig:parafavored_light} The favored regions projected on the planes of 
	$(\Delta m \equiv m_{H^{\pm}} - m_D, m_{H^{\pm}})$ (left panel) and $(m_{W^{\prime}}, \theta_2)$ (right panel) for the light DM mass scenario. The dark (light) orange region represents the $1\sigma$ ($2\sigma$) favored by {\tt CDF-2022}, while the dark (light) gray region indicates the $1\sigma$ ($2\sigma$) favored by the {\tt PDG-2021}.  
	}
\end{figure}

\subsection{Light DM mass scenario}
We show the favored regions from {\tt CDF-2022} and {\tt PDG-2021} for the light DM mass scenario in Fig.~\ref{fig:parafavored_light}. 
Similar to the heavy DM mass scenario, the significant difference between {\tt CDF-2022} and {\tt PDG-2021} 
is the favored region projected on the mass splitting between the charged Higgs and dark Higgs which is shown in the left panel 
of Fig.~\ref{fig:parafavored_light}. 
In particular, within $2\sigma$ region, {\tt CDF-2022} prefers a mass splitting $ 60 \,{\rm GeV} \lesssim \Delta m \lesssim 100$ GeV,  while {\tt PDG-2021} favors a smaller region $ 10 \,{\rm GeV}  \lesssim \Delta m \lesssim 75$ GeV.  
However, unlike the heavy DM mass scenario, both {\tt CDF-2022} and {\tt PDG-2021} in this scenario disfavor the degeneracy of the charged Higgs and dark Higgs masses. 
The favored regions projected on ($m_{W'}-\theta_2$) plane is shown in the right panel of Fig.~\ref{fig:parafavored_light}. 
The results between {\tt CDF-2022} and {\tt PDG-2021} projected on this plane is slightly different. 
 {\tt CDF-2022} prefers a region with the DM mass ($m_{W'} \lesssim 30$ GeV within $2 \sigma$ region) and the mixing angle ($\theta_2 \lesssim 0.15$ rad within $2 \sigma$ region) while {\tt PDG-2021} favors a bit lighter DM mass and higher mixing angle region.
There is a lower bound on the DM mass ($m_{W'} \gtrsim 0.02$ GeV within $2 \sigma$ region) which is 
due to the constraints from the dark photon searches \cite{Ramos:2021omo, Ramos:2021txu}. 
In particular, because the relation between $m_{W'}$ and $m_{Z'}$ shown in~(\ref{MsqZs}), 
the lower bound on $m_{Z'}$ due to NA64 \cite{Banerjee:2018vgk}, E141 \cite{Riordan:1987aw} $\nu$-CAL I \cite{Blumlein:2013cua} experiments can be translated to a lower bound on the DM mass. 
The constraints from the dark photon searches also put a strong upper limit on the gauge couplings $g_H$ and $g_X$  \cite{Ramos:2021omo, Ramos:2021txu}. 
Since the relation between $g_H$ and $\theta_2$ given in~(\ref{eq:gHtheta2relation}),
the upper limit on $g_H$ induces the upper limit on $\theta_2$ as shown in the right panel of Fig.~\ref{fig:parafavored_light}.  

\begin{figure}[tb]
	\includegraphics[width=0.49\textwidth]{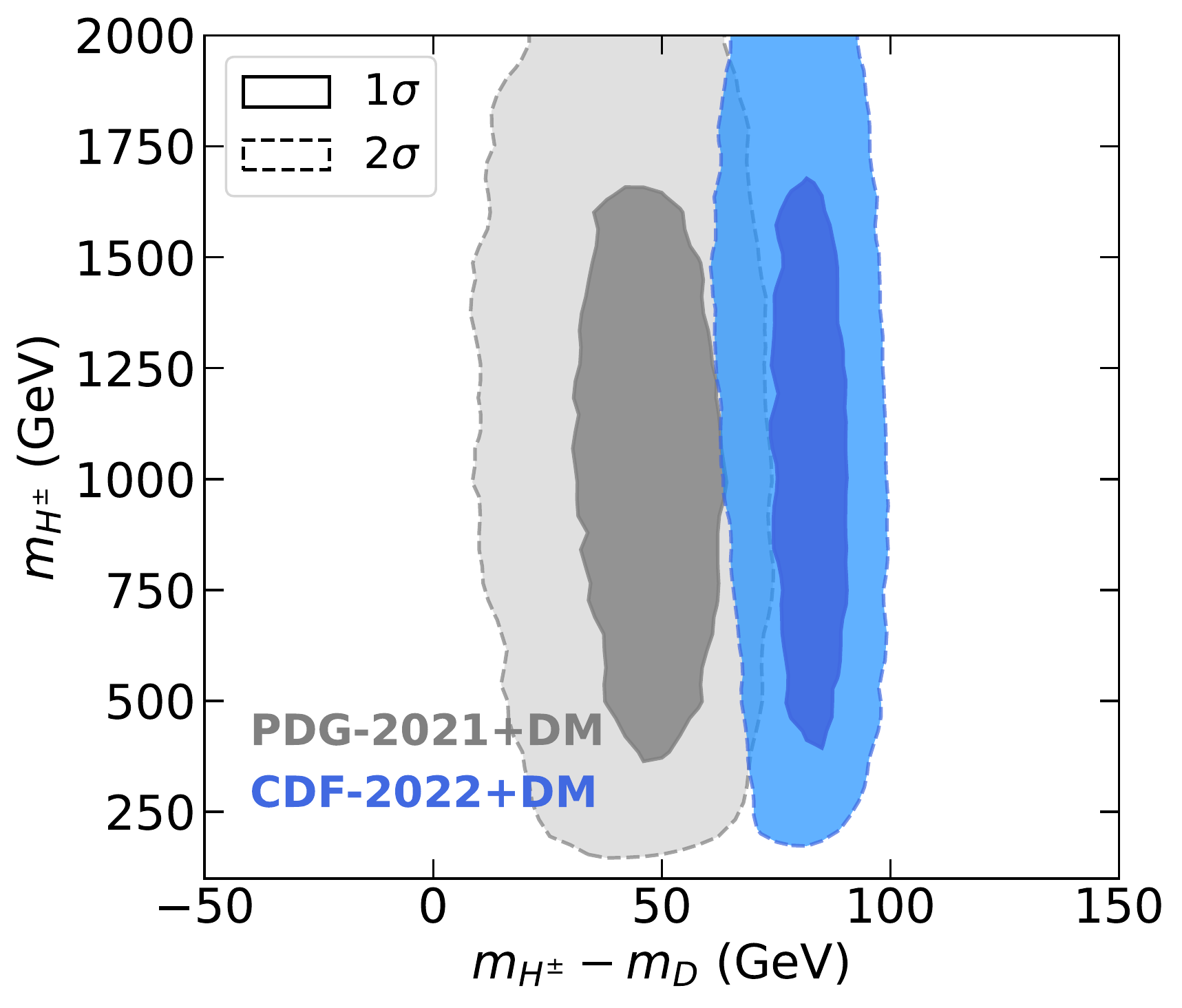}
	\includegraphics[width=0.49\textwidth]{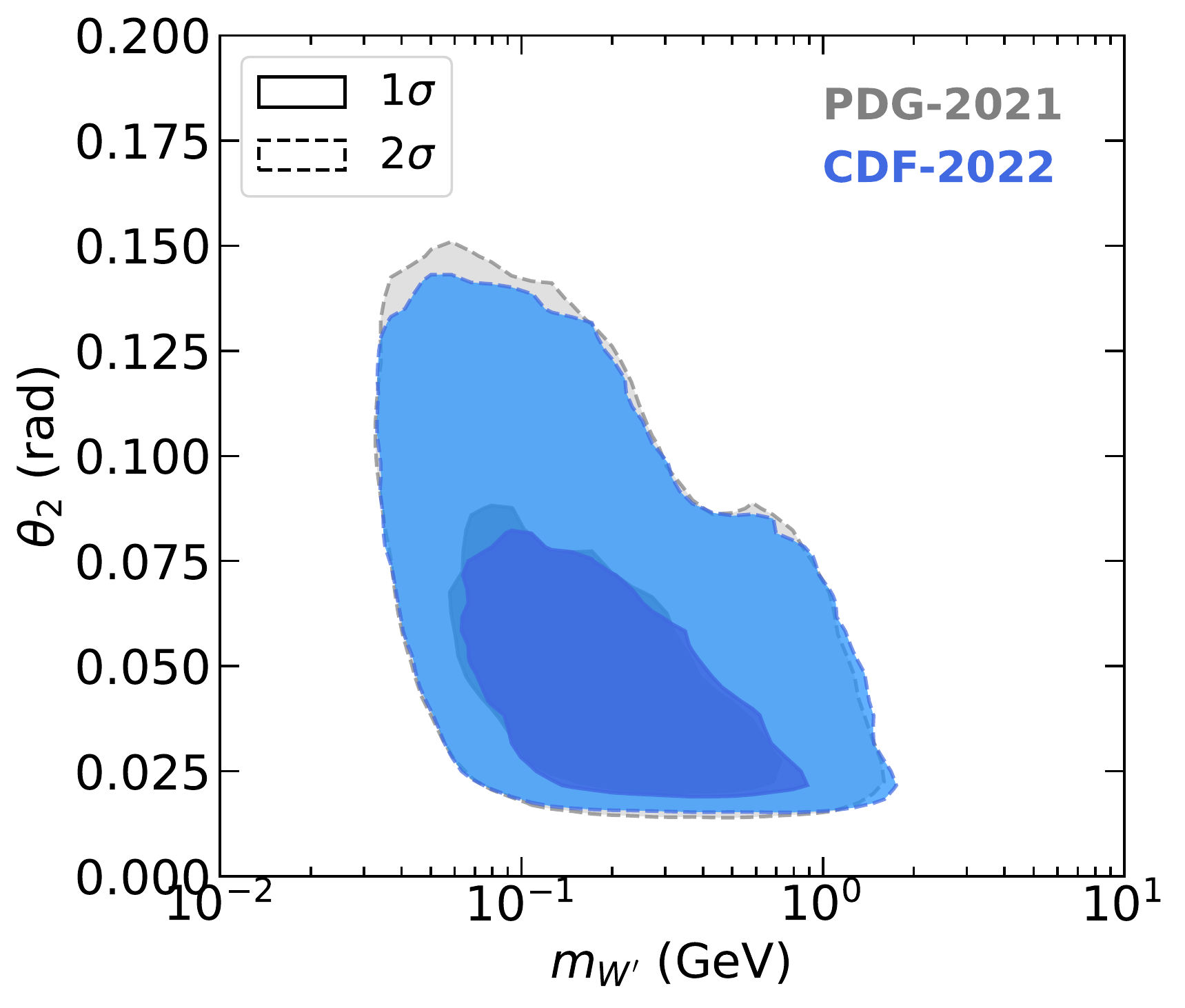}
	\caption{\label{fig:parafavored_lightDM} Same as Fig.~\ref{fig:parafavored_light} for the light DM mass scenario but with the DM constraints included. The dark (light) blue region represents the $1\sigma$ ($2\sigma$) favored by the {\tt CDF-2022+DM}, while the dark (light) gray region indicates the $1\sigma$ ($2\sigma$) favored by the {\tt PDG-2021+DM}.  
	}
\end{figure}

\begin{figure}[tb]
	\includegraphics[width=0.6\textwidth]{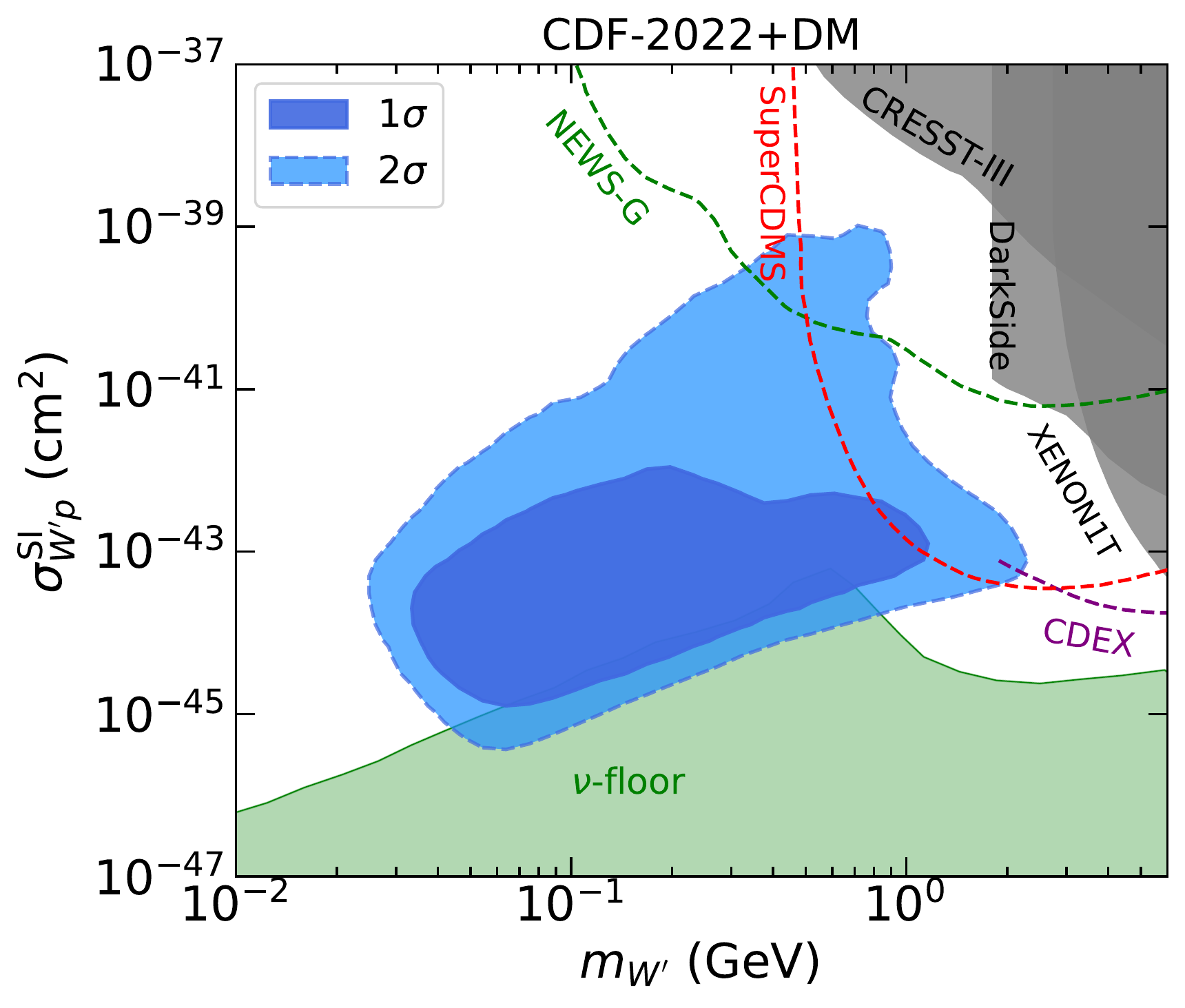}
	\caption{\label{fig:DMxsec_lightmass} The $1\sigma$ (dark blue) and $2\sigma$ (light blue) favored regions of the data from {\tt CDF-2022+DM} projected on the plane of the DM mass and spin independent DM-proton scattering cross section for the light DM mass scenario. The gray regions are the exclusion from CRESST-III~\cite{Angloher:2017sxg}, DarkSide \cite{Agnes:2018ves} and XENON1T \cite{Aprile:2019xxb} experiments. The dashed green, red and purple lines represent the future sensitivities from DM direct detection experiments 
	at NEWS-G \cite{Battaglieri:2017aum}, SuperCDMS \cite{Agnese:2016cpb} and CDEX \cite{Ma:2017nhc}, respectively. Green region is the neutrino floor background.}
\end{figure}

Next, we take into account the DM constraints for this light DM mass scenario.
Due to the scanned mass ranges of $D, H^{\pm}$ and $W'$, the mass splitting between DM and other $h-$parity odd particles $H^{\pm}$ and $D$ are relatively large and hence forbids the DM coannihilation channels. 
The annihilations of $W'^{p} W'^{m}$ to the SM fermion pairs via s-channel with the mediation of $Z', \gamma'$ are dominant processes. 
However the cross sections of such processes are suppressed due to the smallness of the gauge couplings $g_H$ and $g_X$ 
except near the resonant regions where the mass of mediators are about twice DM mass. 
Because the mass of $Z'$ can not satisfy the resonant condition {\it i.e.} $m_{Z'} \not \approx  2 \times m_{W'}$ (see~(\ref{MsqZs})), 
only the annihilation process with a mediator $\gamma'$ near the resonance can provide a DM relic density observed at the Planck Collaboration. 

We show the results with the DM constraints included in Fig.~\ref{fig:parafavored_lightDM}. 
In this scenario, since the DM physics is not significantly impacted by the charged Higgs and dark Higgs,  
the favored regions projected on ($\Delta m$, $m_{H^{\pm}}$) plane as shown in the left panel of Fig.~\ref{fig:parafavored_lightDM} are almost unchanged as compared with the one without DM constraints as shown in the left panel of Fig.~\ref{fig:parafavored_light}. 
On the other hand, due to the DM direct detection constraints from CRESST-III~\cite{Angloher:2017sxg}, DarkSide \cite{Agnes:2018ves} and XENON1T \cite{Aprile:2019xxb} experiments, 
the DM mass is required to be $m_{W'} \lesssim 2$ within $2\sigma$ region as shown in the right panel of Fig.~\ref{fig:parafavored_light}. 
In the same panel, we also see that it is required to have a lower bound on $\theta_2 \gtrsim 4 \times 10^{-3}$ rad. This is because of the DM relic density observed at the Planck Collaboration. A smaller $\theta_2$ results in a smaller $g_H v/{m_{W'}}$ value as shown in (\ref{eq:gHtheta2relation}), 
which can cause a smaller DM annihilation cross section and eventually lead to an overabundant DM in the universe.

In Fig.~\ref{fig:DMxsec_lightmass}, we project the favored regions from the {\tt CDF-2022+DM} data on the plane of the DM mass and spin independent DM-proton scattering cross section. 
The current constraints from CRESST-III~\cite{Angloher:2017sxg}, DarkSide \cite{Agnes:2018ves} and XENON1T \cite{Aprile:2019xxb} experiments are shown as the gray shaded regions. These constraints put an upper limit on the DM mass in this scenario.  
We find out that a portion of the favored region at $m_{W'} \sim 1$ GeV and $\sigma_{W'p}^{\rm SI} \sim
[10^{-44} - 10^{-39}] \, {\rm cm}^2$ can be probed by future DM direct detection experiments 
at NEWS-G \cite{Battaglieri:2017aum}, SuperCDMS \cite{Agnese:2016cpb} and CDEX \cite{Ma:2017nhc}. 
Note also a small fraction of the $2 \sigma$ favored region lies below the neutrino floor background (light green).

\subsection{The processes $h \to \gamma \gamma$ and $h \to Z \gamma$ at the LHC}

Given the favored parameter space obtained in previous two subsections for the heavy and light DM scenarios, one would like to study their impacts 
to the diphoton and $Z \gamma$ channels from the Higgs decays and the future collider searches for these processes at the hadron collider.
A precise measurement of the signal strength of these processes at the colliders could
reveal the existence of new particles coupled to the Higgs (see e.g. \cite{Chiang:2012qz, Carena:2012xa}). 
For the diphoton channel, 
measurements with the current $13$ TeV LHC data yield a signal strength 
via the gluon-gluon fusion production mechanism: $\mu^{\rm ggh}_{\gamma\gamma} = 0.96 \pm 0.14$ \cite{Aad:2019mbh}. 
For $14$ TeV LHC with $3\,{\rm ab}^{-1}$ luminosity, the signal strength is expected to be measured with a $\pm 0.04$ uncertainty \cite{Cepeda:2019klc}. 
On the other hand, the $Z \gamma$ channel has not yet been observed with the current data at the LHC \cite{ATLAS:2017zdf, CMS:2018myz}. With an expected uncertainty of $24\%$ for the signal strength measurement at the $14$ TeV LHC with $3\,{\rm ab}^{-1}$ luminosity, this $Z\gamma$ channel can be observed with 4.9 $\sigma$ significance at ATLAS experiment \cite{Cepeda:2019klc}. 

The analytical expressions for the one-loop amplitudes of $h \to \gamma \gamma$ and $h \to Z \gamma$ 
in the G2HDM are given in the Appendix. 
As compared with the SM prediction, the production rate of these two channels can be modified due to new contributions from the charged Higgs and new charged heavy fermions running inside the loops as well as effects from the mixing angle $\theta_1$ between the SM Higgs boson $h_{\rm SM}$ 
and the hidden scalar $\phi_H$ in the model. We note that in this analysis, we assume all new heavy fermions to be degenerated and fixed their masses to be $3$ TeV. 
We select the data points from {\tt CDF-2022+DM} with $\Delta \chi^2 = \chi^2 - \chi^2_{\rm min} < 5.99$ 
where $\chi^2$ is the total $\chi^2$ calculated from the Higgs data at the LHC, $Z$-boson mass measurement from LEP II, oblique parameters from Ref. \cite{deBlas:2022hdk}, DM relic density from Planck collaboration and the one-side limits from $Z'$ searches, DM direct detection and Higgs invisible decay. 

\begin{figure}[tb]
	\includegraphics[width=0.45\textwidth]{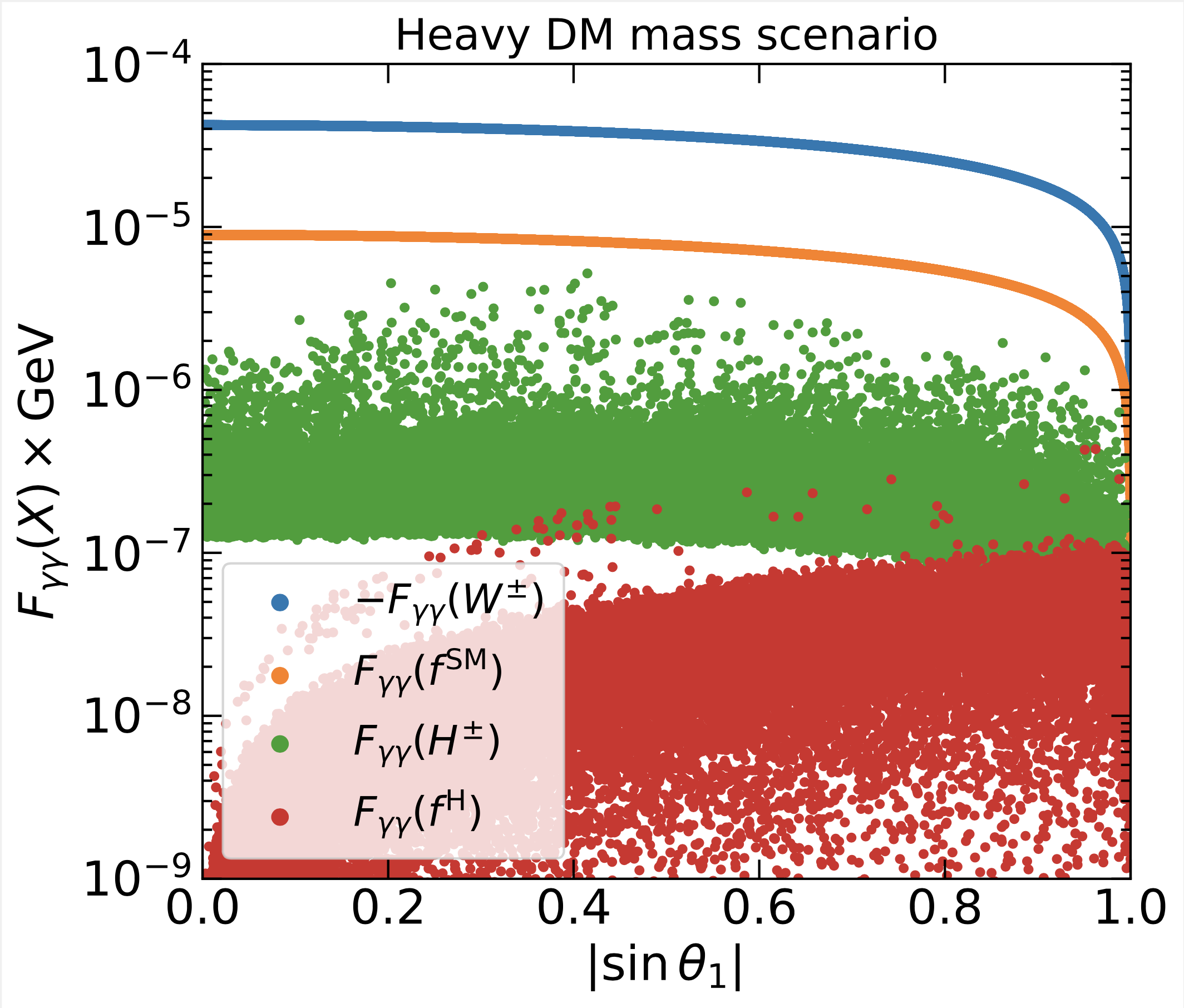}
	\includegraphics[width=0.45\textwidth]{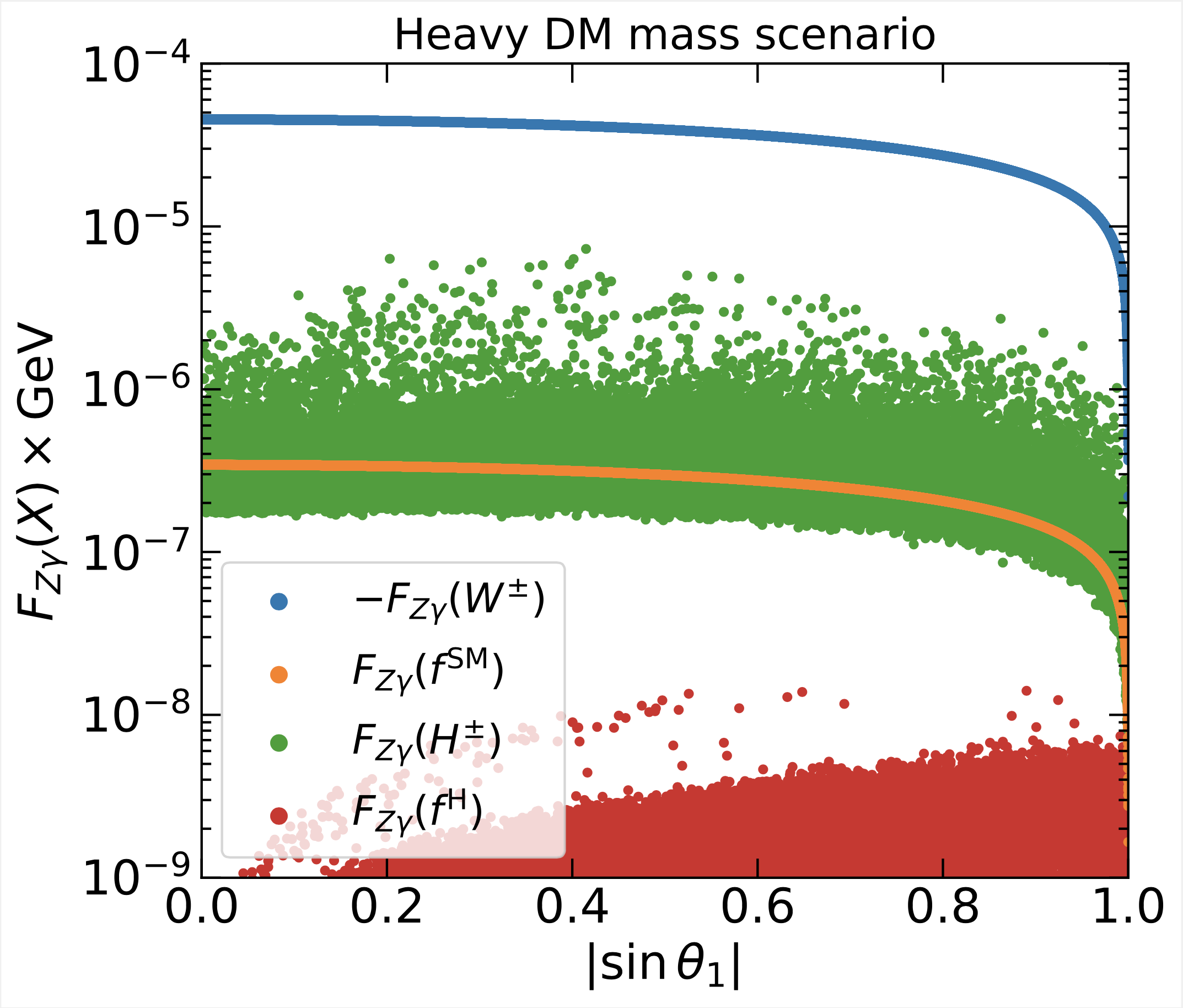}
	\caption{\label{fig:Fij_heavy} The  one-loop  form factors for the individual particle running inside the loop of $h \to \gamma \gamma$ (left panel) and $h \to Z \gamma$ (right panel) processes
	as a function of $|\sin \theta_1|$. The data points are taken in $\Delta \chi^2 < 5.99$ region from {\tt CDF-2022+DM}  for the heavy DM mass scenario. The orange, green and red points represent the form factors for the SM fermions, charged Higgs, new heavy fermions respectively. The blue points represent the negative value of form factor for the $W$ boson. Note that $F_{\gamma \gamma}(X) \equiv F_{1} (X)$ defined in (\ref{FWloopgg}, \ref{FfSMloopgg}, \ref{FfHloopgg}, \ref{FCHloopgg}) and $F_{Z \gamma}(X) \equiv F_{11} (X)$ defined in (\ref{FWloopZg}, \ref{FfSMloopZg}, \ref{FfHloopZg}, \ref{FCHloopZg}) in the Appendix. Here $X \equiv (W^\pm, f^{\rm SM}, f^H, H^{\pm})$.
	}
\end{figure}

\begin{figure}[tb]
	\includegraphics[width=0.45\textwidth]{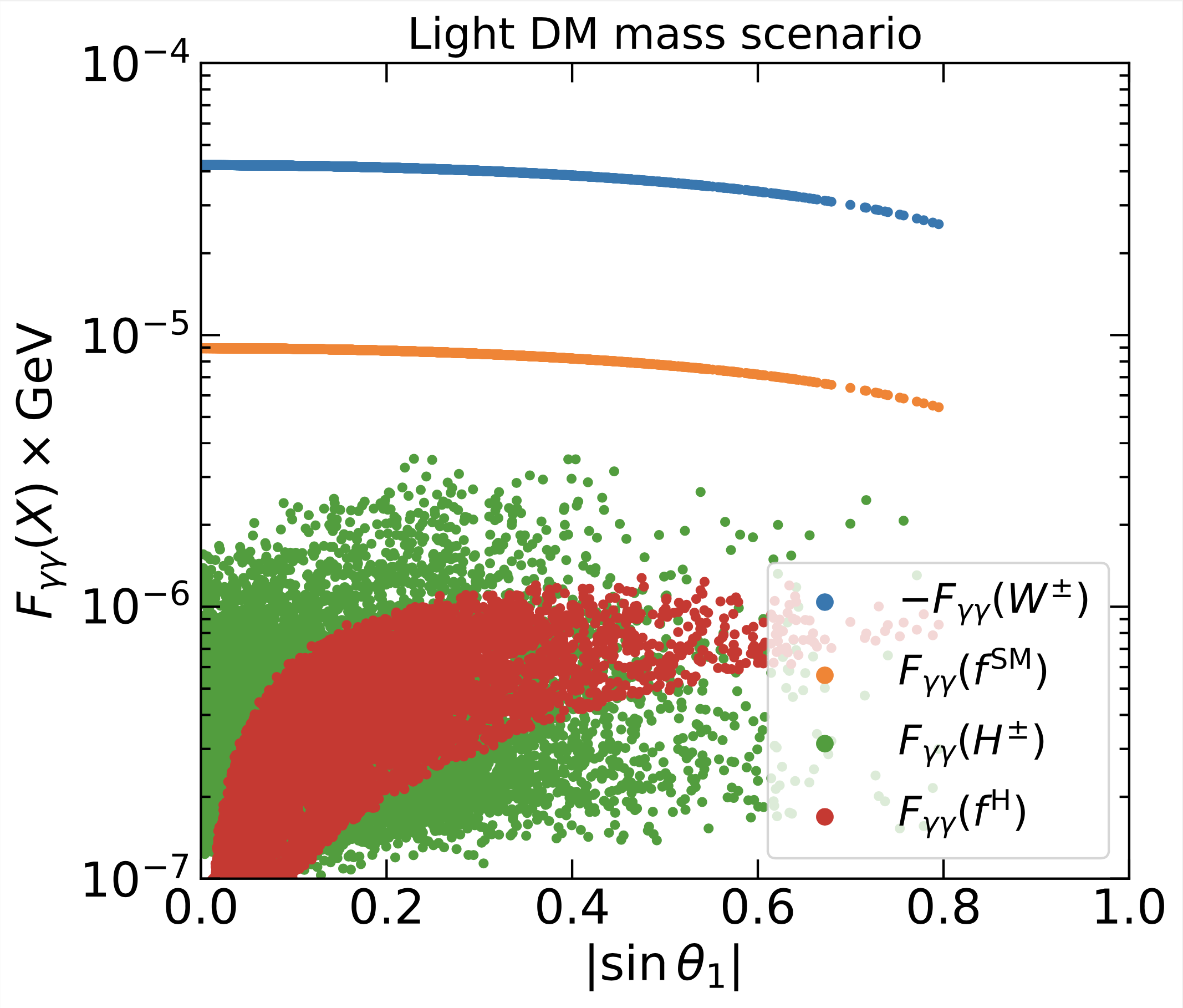}
	\includegraphics[width=0.45\textwidth]{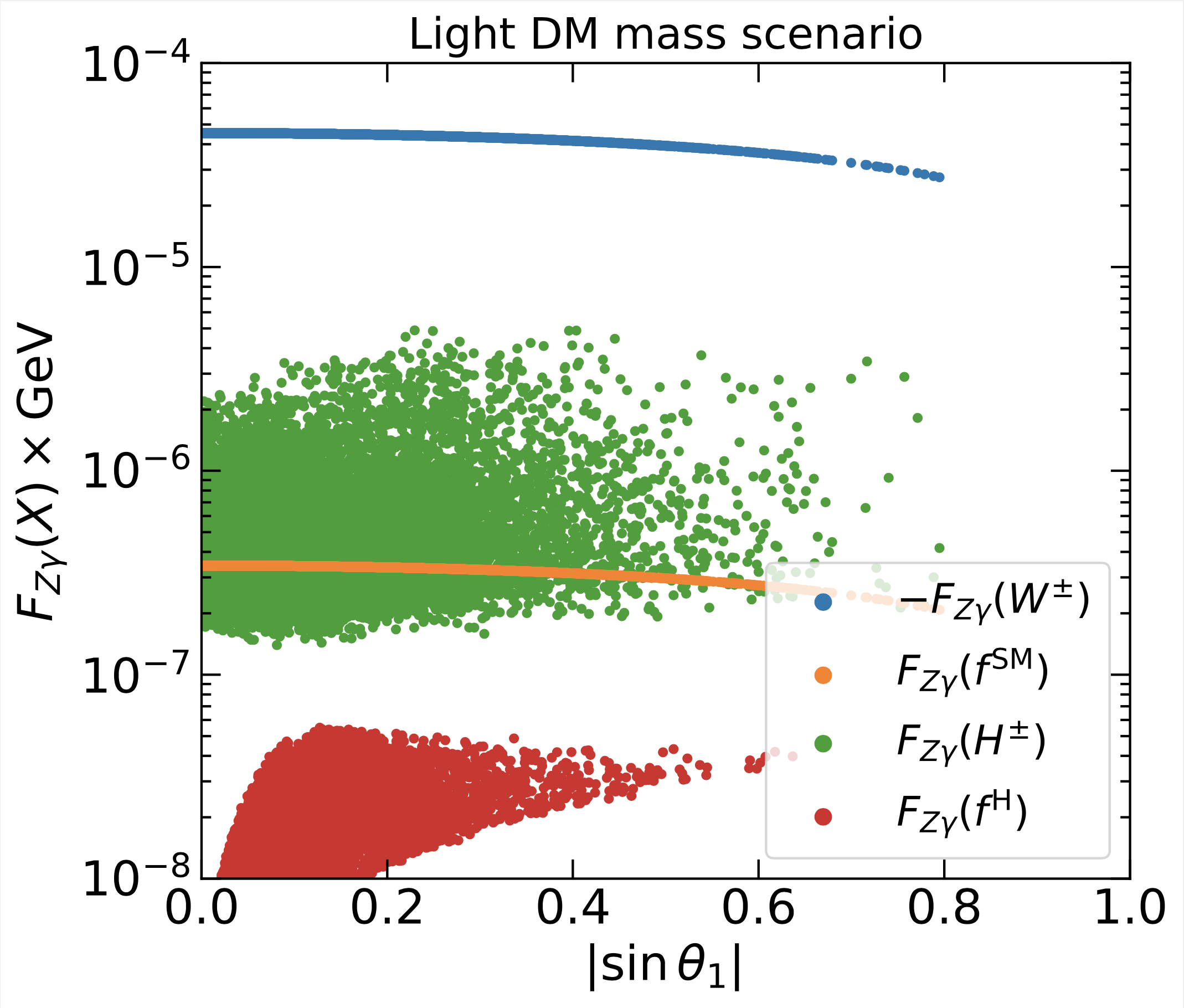}
	\caption{\label{fig:Fij_light} Same as Fig.~\ref{fig:Fij_heavy} but for the light DM mass scenario.
	}
\end{figure}

Fig.~\ref{fig:Fij_heavy} shows the  one-loop  form factor for the individual particle including $W^{\pm}$, $H^\pm$, charged SM fermions (mainly  top quark) and new charged
heavy fermions running inside the loop of $h \to \gamma \gamma$ (left panel) and $h \to Z \gamma$ (right panel) processes as a function of $|\sin \theta_1|$ 
in the heavy DM scenario.
In both $h \to \gamma \gamma$ and $h \to Z \gamma$ processes, 
the $W^\pm$ form factor gets negative values
but its magnitude is dominant. 
From the right panel of Fig.~\ref{fig:DMrelicparafavored} and the left panel of Fig.~\ref{fig:parafavored_lightDM}, 
we learnt that our scan results prefer a heavy charged Higgs mass in both heavy and light DM mass scenarios.
One expects the charged Higgs form factor is rather small. Indeed, in the small mixing angle region, $F_{\gamma \gamma, Z\gamma} (H^{\pm}) \sim {\cal O} (10^{-7} - 10^{-5})/$ GeV  and it would be further suppressed in the large mixing angle region.
The charged Higgs form factor is smaller than the SM top quark form factor in the $h \to \gamma \gamma$ process but it can be larger in the $h \to Z \gamma$ process. 
We also see that the new charged heavy fermions are suppressed due to the large VEV $v_\Phi$ in this data. 
We note that the SM fermion and new heavy fermion form factors for the $h \to \gamma \gamma$ process are about one order of magnitude larger 
than those for the $h \to Z \gamma$ process, 
while the $W$ boson (as well as the charged Higgs) form factors stay almost the same for these two processes. 
In the limit of no mixing between the SM Higgs boson $h_{\rm SM}$ and the hidden scalar $\phi_H$, {\it i.e.} $\sin \theta_1 = 0$, 
the new heavy fermion form factor would be vanished. 

Fig.~\ref{fig:Fij_light} shows a similar plot as Fig.~\ref{fig:Fij_heavy} but for the light DM mass scenario. 
In this scenario, $m_{h_1} > 2 \,m_{W'}$, the Higgs boson can decay invisibly into a pair of $W^{\prime (p,m)}$. 
The current data from ATLAS \cite{ATLAS:2020cjb} can put an upper limit on the mixing angle $\theta_1$. 
In particular it requires $|\sin \theta_1| \lesssim 0.8$. 
Furthermore, a lighter DM mass also results in a smaller $v_\Phi$ value, this would enhance the contribution of new heavy fermions in the loop. 
As shown in the left panel of Fig.~\ref{fig:Fij_light} for the $\gamma\gamma$ case,
the contribution from the new heavy fermions can be comparable with the charged Higgs when $|\sin \theta_1| \gtrsim 0.1$.

\begin{figure}[tb]
	\includegraphics[width=0.45\textwidth]{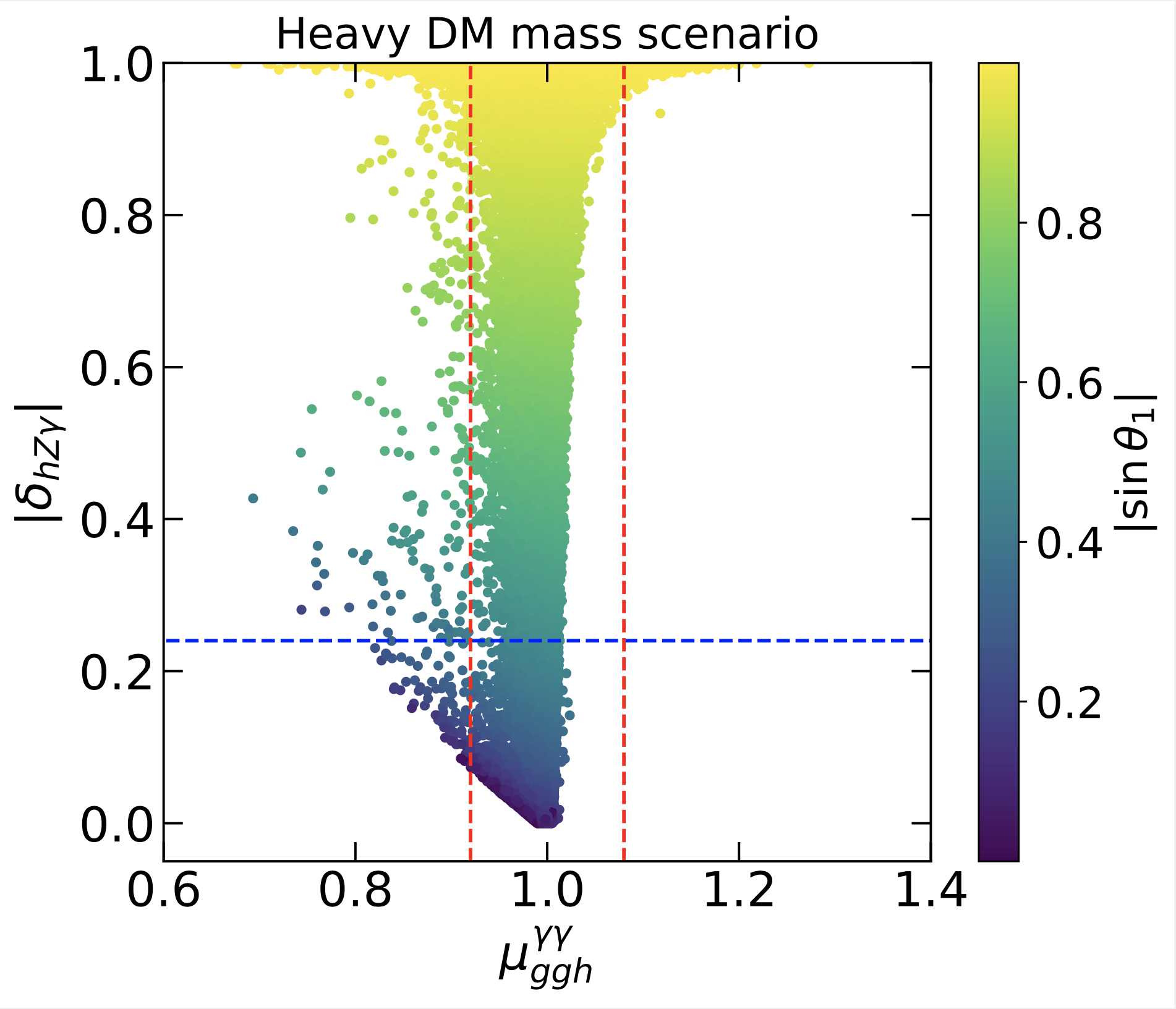}
	\includegraphics[width=0.45\textwidth]{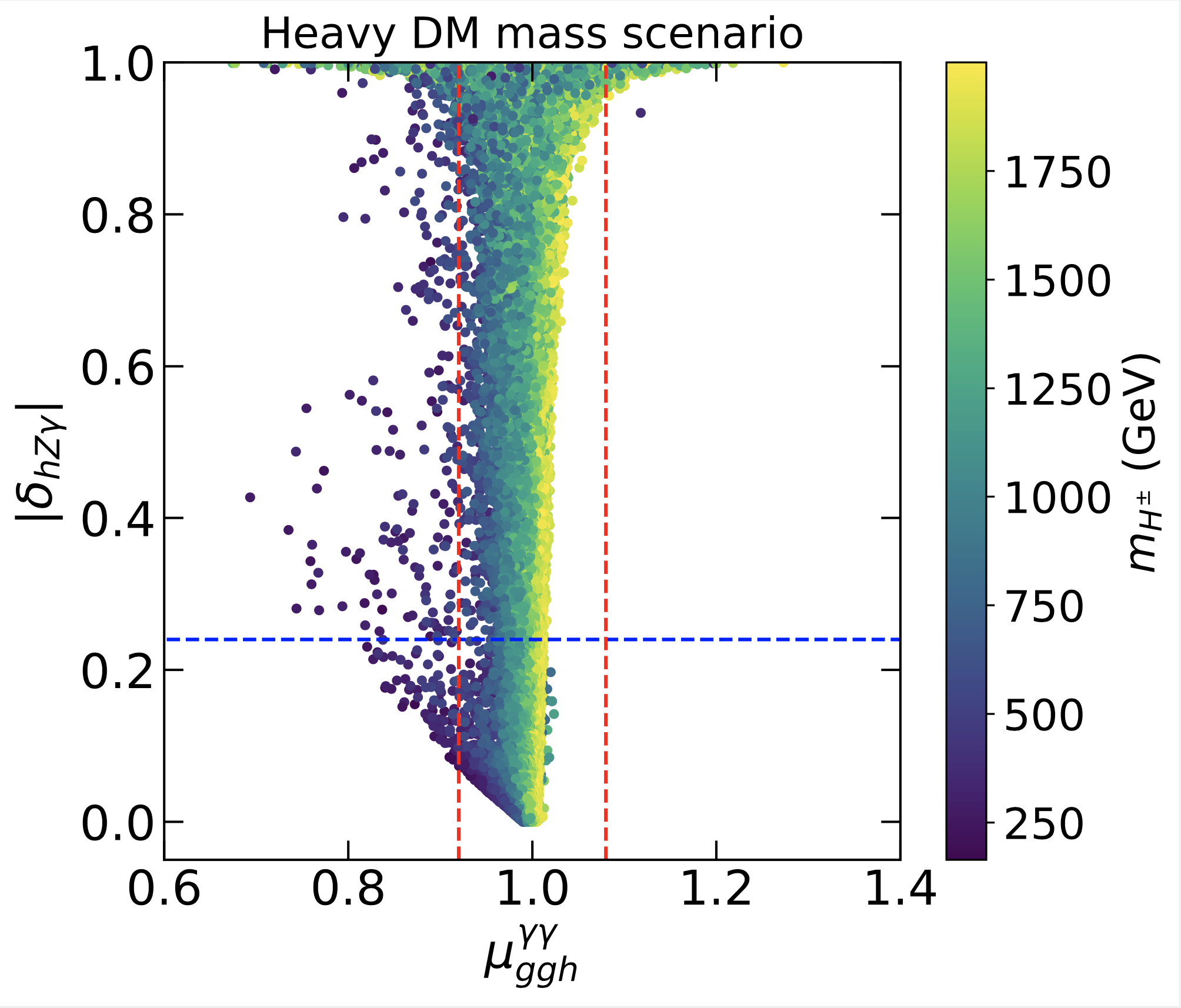}
	\caption{\label{fig:futurecollider_heavy} The favored data points from {\tt CDF-2022+DM} for the heavy DM mass scenario projected on the ($\mu^{\rm ggh}_{\gamma\gamma}, |\delta_{hZ\gamma}|$) plane. In the left panel the color gradient represents the mixing angle $|\sin {\theta_1}|$ while on the right panel it represents the charged Higgs mass $m_{H^\pm}$. The region between the two vertical red lines is $2\sigma$ region of the expected diphoton signal strength measurement with a $4\%$ uncertainty at the HL-LHC \cite{Cepeda:2019klc}. The horizontal dashed blue line represents the expected measurement precision of $24\%$ for $Z \gamma$  production at the HL-LHC \cite{Cepeda:2019klc}.}
\end{figure}

\begin{figure}[tb]
	\includegraphics[width=0.45\textwidth]{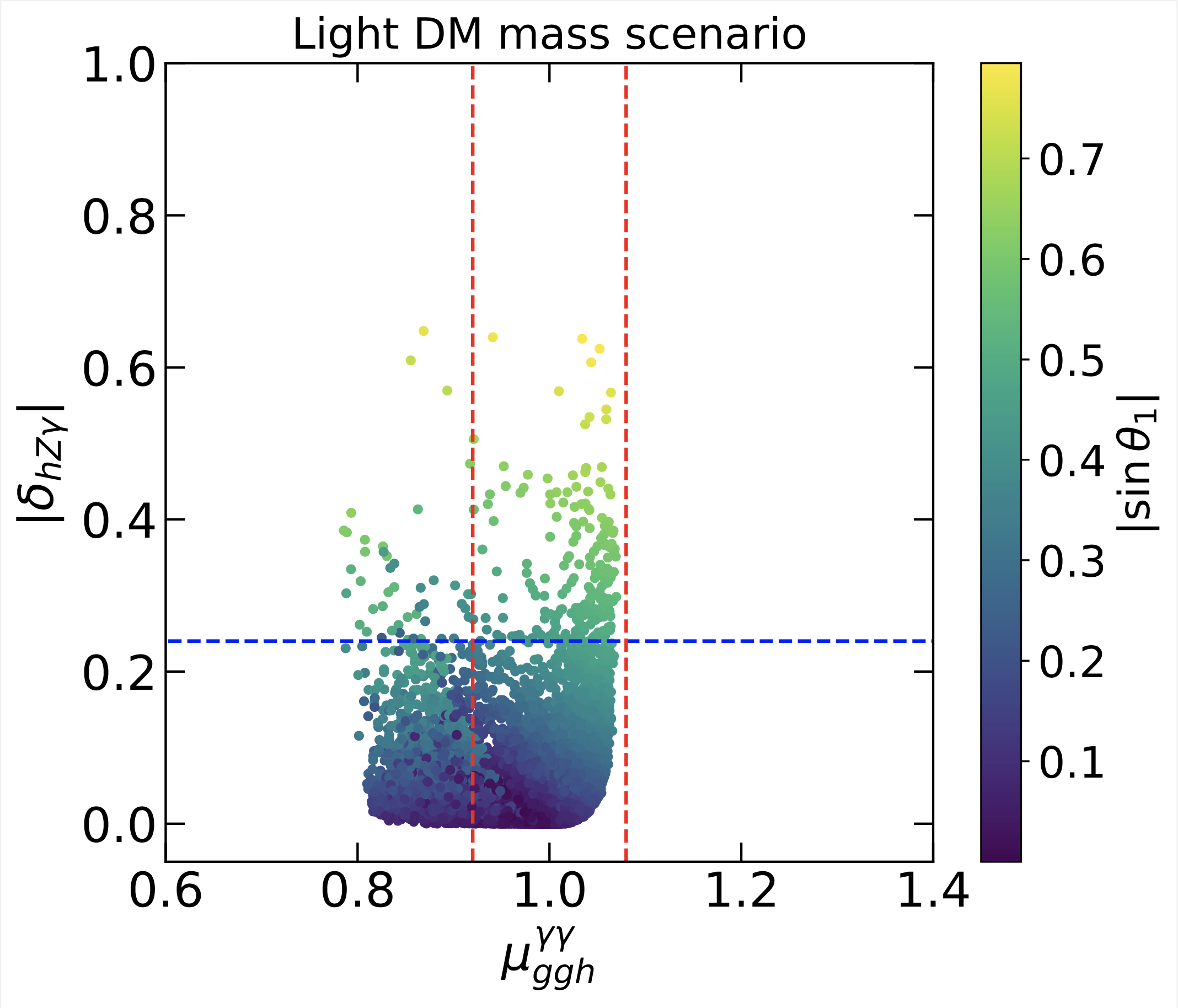}
	\includegraphics[width=0.45\textwidth]{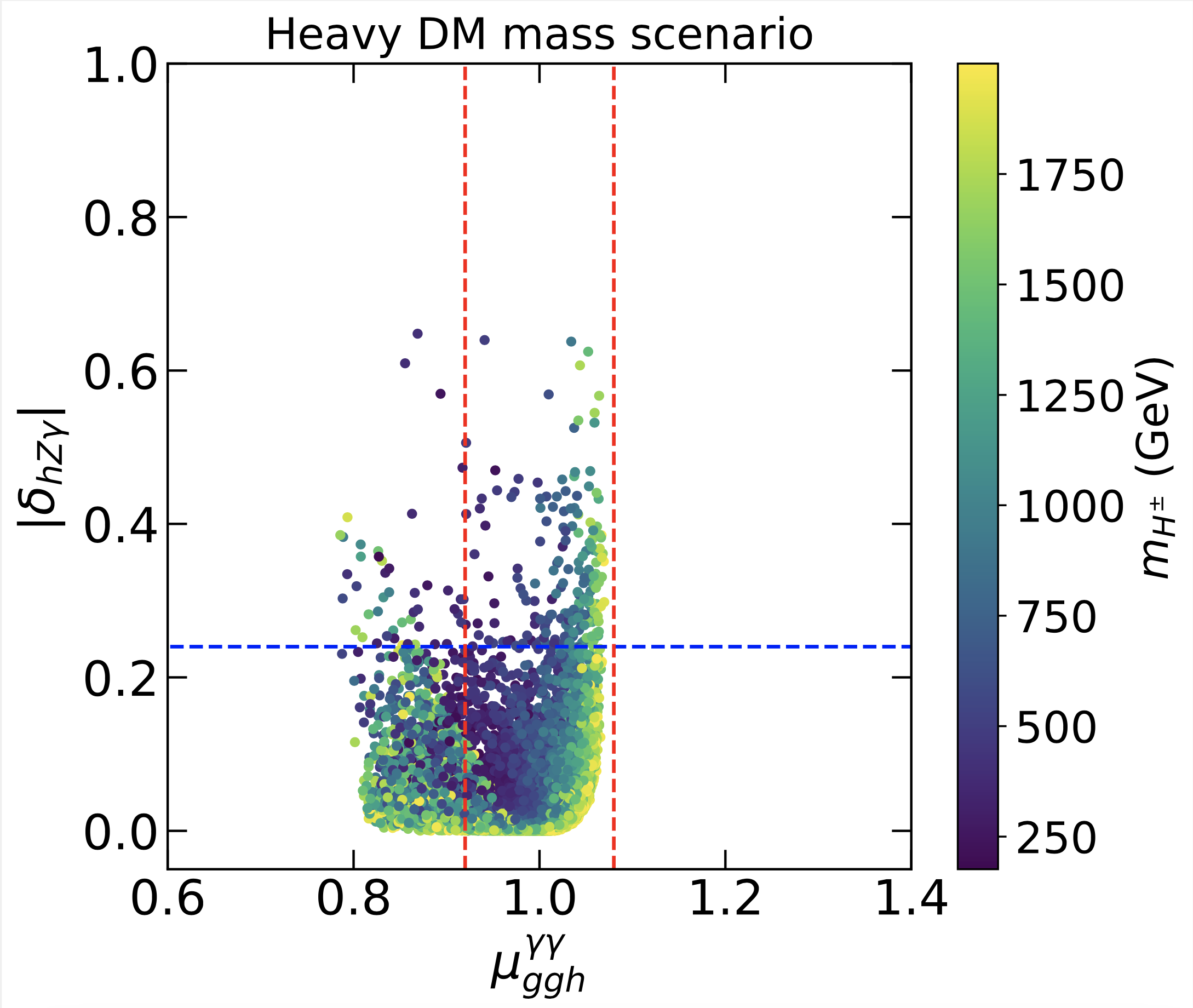}
	\caption{\label{fig:futurecollider_light} Same as Fig.~\ref{fig:futurecollider_heavy} but for the light DM mass scenario.}
\end{figure}

In Fig.~\ref{fig:futurecollider_heavy}, we project the favored data points from {\tt CDF-2022+DM} for the heavy DM mass scenario on the 
($\mu^{\rm ggh}_{\gamma\gamma}, |\delta_{hZ\gamma}|$) plane, 
where 
\beq
\delta_{hZ\gamma} = 1 - \frac{\Gamma ({h \to Z \gamma)}} { \Gamma ({h_{\rm SM} \to Z_{\rm SM} \gamma}) } 
\eeq
is the normalized deviation of the $h \to Z\gamma$ decay width from its SM value. 
The signal strength of diphoton production from the gluon-gluon fusion is given by 
\be
\mu_{\gamma\gamma}^{\rm ggh} =  \frac{\Gamma_{h_{\rm SM}}} {\Gamma_{h}} 
\frac{\Gamma ({h \to gg}) \Gamma ({h \to \gamma\gamma}) } 
{\Gamma ({h_{\rm SM} \to gg}) \Gamma ( {h_{\rm SM} \to \gamma\gamma} )} \; ,
\ee 
where the decay width of $h$ into two gluons in the model can be found in~\cite{Huang:2015wts}. 

The deviation $|\delta_{hZ\gamma}|$ can be large if the mixing angle $\theta_1$ is large as shown in the left panel of Fig.~\ref{fig:futurecollider_heavy}. Similarly, $\mu^{\rm ggh}_{\gamma\gamma}$ can be also impacted by this mixing effects, a larger mixing angle $\theta_1$ results in a wider region of $\mu^{\rm ggh}_{\gamma\gamma}$. 
Moreover, due to the destructive interference between the charged Higgs and $W$ boson contributions, the diphoton signal strength becomes smaller when the charged Higgs has a lighter mass as shown in the right panel of Fig.~\ref{fig:futurecollider_heavy}.
Future measurements for the diphoton and $Z\gamma$ production from the HL-LHC can probe a large portion of the {\tt CDF-2022+DM} favored region in this heavy DM scenario. In particular, the large mixing region $|\theta_1| > 0.5$ can be excluded by the expected measurement precision of $24\%$ for $Z \gamma$  production at the HL-LHC \cite{Cepeda:2019klc} which is shown by the horizontal dashed blue line in Fig.~\ref{fig:futurecollider_heavy}. The low mass region of charged Higgs can be probed 
by the expected diphoton signal strength measurement with a $4\%$ uncertainty at the HL-LHC \cite{Cepeda:2019klc}.  

Similar plots as Fig.~\ref{fig:futurecollider_heavy} but for the light DM mass scenario are shown in Fig.~\ref{fig:futurecollider_light}. 
Both $|\delta_{hZ\gamma}|$ and $\mu^{\rm ggh}_{\gamma\gamma}$ can be impacted by the mixing effects. However, due to the constraint from the Higgs invisible decays on the mixing angle $\theta_1$, the deviations on diphoton and $Z\gamma$ decay widths from the SM prediction are less significant as compared with the heavy DM mass scenario. 
The dependence of $|\delta_{hZ\gamma}|$ and $\mu^{\rm ggh}_{\gamma\gamma}$ on the charged Higgs mass is also less significant as shown in the right panel of Fig.~\ref{fig:futurecollider_light}.
We note that, in this scenario, Higgs boson is kinematically allowed to decay into light neutral gauge bosons $Z'Z'$, $\gamma '\gamma'$, $ZZ'$ and $Z \gamma'$, however the branching ratio of these processes are minuscule due to the smallness of the gauge couplings $g_H$ and $g_X$. 
As also shown in Fig.~\ref{fig:futurecollider_light}, precision measurements on the diphoton and $Z\gamma$ productions at the HL-LHC can further constraint the parameter space of the model especially on the mixing angle $\theta_1$ for the light DM mass scenario.

\section{Conclusions}
\label{sec6}

Motivated by the new high-precision measurement of the SM $W$ boson mass by the CDF II~\cite{CDF:2022hxs}, 
we scrutinize the minimal G2HDM by computing the contributions from the inert Higgs doublet, hidden Higgs doublet, hidden neutral dark gauge bosons and extra heavy fermions
in the model to the oblique parameters $S$, $T$, and $U$. 
We found out that only inert Higgs doublet can give a significant contribution while contributions from other new particles in the model are minuscule. 

The favored regions on the model parameter space are obtained using the updated electroweak global fits of these oblique parameters~\cite{deBlas:2021wap,deBlas:2022hdk} 
and the current constraints in the model including the theoretical constraints on the scalar potential, the Higgs data from the LHC,
$Z'$, dark photon physics constraints as well as relic density, direct detection and Higgs invisible decays constraints for the dark matter candidate $W'$. 
We initiated  two scenarios based on the dark matter mass, 
one is heavy DM mass scenario and another is light DM mass scenario. 
By comparing with the old global fit values for the oblique parameters from PDG, 
we found out that the CDF $W$ boson mass measurement significantly impacts 
the mass splitting $\Delta m = (m_H^\pm - m_D)$ in the inert $h$-parity odd Higgs doublet $H_2$. 
In particular, the result using the updated global fit values~\cite{deBlas:2022hdk} favors a larger mass splitting ($72 \,{\rm GeV} \lesssim \Delta m \lesssim  118 \,{\rm GeV}$ for the heavy DM mass scenario and  $60 \,{\rm GeV} \lesssim \Delta m \lesssim  100 \,{\rm GeV}$ for the light DM mass scenario) as compared with one using the old global fit values ($33 \,{\rm GeV} \lesssim \Delta m \lesssim  87 \,{\rm GeV}$ for the heavy DM mass scenario and  $10 \,{\rm GeV} \lesssim \Delta m \lesssim  75 \,{\rm GeV}$ for the light DM mass scenario) 
from the PDG.
We also found that to satisfy the DM relic density observed at the Planck collaboration, 
it is required the $W'$ coannihilation processes to be occurred for the heavy DM mass scenario and 
the nearly resonant annihilation process with the dark photon as a mediator for the light DM mass scenario.
The former requires the differences $(m_{H^\pm} - m_{W^{\prime}})/{m_{W^{\prime}}}$ and $(m_{D} - m_{W^{\prime}})/{m_{W^{\prime}}}$ to be in $\sim {\cal O}(10^{-2}, 10^{-1})$, while the latter requires the mediator mass to be around twice the DM mass.
For the heavy DM mass scenario, the favored dark matter mass is around $1$ TeV and can be probed by future dark matter 
direct detection experiments at DarkSide-20k and DARWIN as shown in Fig.~\ref{fig:DMxsec}. 
Whereas, for the light DM mass scenario, the favored DM mass locates at sub-GeV $\sim {\cal O}(0.1-1)$ GeV and also 
can be probed by future dark matter 
direct detections at CDEX, NEWS-G and SuperCDMS as shown in Fig.~\ref{fig:DMxsec_lightmass}.

Furthermore we studied the diphoton and $Z \gamma$ productions from the Higgs decays and the detectability of these processes at the LH-LHC. We showed that the contribution from the charged Higgs can be larger than the SM fermions (mainly top quark) for the $h \to Z \gamma$ process but not for the diphoton channel. In the low charged Higgs mass region, the contribution from charged Higgs can be significant and thus reduces the decay widths of 
$h \to \gamma \gamma$ and $h \to Z \gamma$ due to the destructive interference with the $W$ boson contribution. 
We also showed that the deviations of the diphoton and $Z \gamma$ productions from the SM predictions are 
highly depending on the mixing angle $\theta_1$ between the SM Higgs boson and the scalar from the hidden doublet. 
A large portion of the favored region, especially the large mixing angle $\theta_1$ and low charged Higgs mass regions, 
can be probed by more precise measurement of the $h \to \gamma \gamma$ signal strength and the detection of $h \to Z \gamma$ process at the HL-LHC. 

Other rare decay modes 
$h \to \gamma \gamma^\prime, \gamma Z^\prime$ (one-loop) and $h \to \gamma^\prime \gamma^\prime, Z^\prime Z^\prime$, and $\gamma^\prime Z^\prime$ (tree)
in G2HDM may be kinematical allowed and thus have important impacts at collider physics. 
For instance, the dark photon ($\gamma^\prime$) and dark $Z$ ($Z^\prime$) may be long-lived, traversing some macroscopic distances 
for small enough $g_X$ and $g_H$ before they decay into SM light fermion pairs. We reserve the exploration of these modes for future studies. 
For a recent analysis of $h \to \gamma \gamma^\prime$ at the LHC, see~\cite{Beauchesne:2022fet}.

Before closing, we reiterate that the results of the present work are obtained under the approximation of the new gauge couplings
of $SU(2)_H \times U(1)_X$ are much smaller than the SM gauge couplings, as suggested from the analysis of $Z$ mass shift and 
dark matter direct detection in our previous works~\cite{Ramos:2021txu,Ramos:2021omo}.
Consequently we can ignore the loop contributions from the extra gauge bosons $(W^{\prime\, (p,m)}, Z^\prime, \gamma^\prime$) to the oblique parameters 
since the effects would be down by the loop factor $1/16\pi^2$ compared with the tree level mass mixing effects studied in subsection \ref{subsec4a}.
Furthermore, in analogous to the SM case, there should be 8 more vacuum polarization amplitudes 
$i \Pi^{\mu\nu}_{W^{\prime \, p} W^{\prime \, m}}$, 
$i \Pi^{\mu\nu}_{Z^\prime Z^\prime}$, 
$i \Pi^{\mu\nu}_{\gamma^\prime \gamma^\prime}$,
$i \Pi^{\mu\nu}_{Z^\prime \gamma^\prime}$,
$i \Pi^{\mu\nu}_{\gamma \gamma^\prime}$,
$i \Pi^{\mu\nu}_{\gamma Z^\prime}$,
$i \Pi^{\mu\nu}_{Z \gamma^\prime}$, and 
$i \Pi^{\mu\nu}_{Z Z^\prime}$  
that are needed to be worried about. The formalism of the oblique parameters by Peskin and Takeuchi~\cite{Peskin:1991sw}
would have to extend properly so as to take into account all new hidden particle contributions to all possible oblique parameters in G2HDM. 
Such a task is very interesting but beyond the scope of this present work.
We hope to return to this challenge in the future.

\clearpage

\section*{Acknowledgments}

We thank Dr. Raymundo Ramos for sharing his computer codes from our previous collaborations. 
The analysis presented here was done using the computation resources at the Tsung Dao Lee Institute, Shanghai Jiao Tong University.
This work is supported in part by the National Science and Technology Council (NSTC) of Taiwan under Grant Nos. 110-2112-M-001-046, 111-2112-M-001-035  
(TCY) and by National Natural Science Foundation of China under Grant No. 19Z103010239 (VQT).
VQT would like to thank the High Energy Theory Group at the Institute of Physics, Academia Sinica, Taiwan for its hospitality.

\clearpage

\section*{Appendix}
\label{appendix}

In this Appendix we provide the corrected general analytical expressions for the one-loop amplitudes of the two processes 
$h_i \to \gamma \gamma$ and $h_i \to Z_j \gamma$~\footnote{See for example Refs.~\cite{Gunion:1989we,Gamberini:1987sv,Weiler:1988xn,Chen:2013vi,Hue:2017cph}
for the computation of this process in a variety of BSM.} in G2HDM.
As mentioned in the text, the $h_1$ and $Z_1$ are identified as the observed $125.38 \pm 0.14$ GeV~\cite{CMS:2020xrn} Higgs 
$(h)$ and $91.1876 \pm 0.0021$  GeV~\cite{Zyla:2020zbs}  neutral gauge boson $(Z)$ respectively. 
Under the assumption that $g_H, g_X \ll g, g^\prime$ used in the present analysis, 
one can set the couplings $g_H$ and $g_X$ to be zero and the mixing matrix element ${\mathcal O}^G_{ij} = \delta_{ij}$ 
in the following formulas. However the analytical expressions are valid for general mixing cases.

\noindent
\underline{$h_i \to Z_j \gamma$}

The decay rate for $h_i \to Z_j  \gamma $ is 
\be
\Gamma( h_i \to Z_j \gamma ) = \frac{1}{32 \pi} m_{h_i}^3 \left( 1 - \frac{m_{Z_j}^2}{m_{h_i}^2} \right)^3 
\biggl\vert F^1_{i j }  + F^{1/2}_{i j }  + F^0_{i j }  \biggr\vert^2 \; ,
\label{RatehiZg}
\ee
where $F^s_{ij}$ with $s=0,1/2,1$ denotes the loop form factor for charge particle with spin equals $0,1/2,1$ respectively running inside the loop.

In G2HDM, the only charged spin 1 particle is the SM $W^\pm$, thus $F^1_{i j } = F_{i j } (W^\pm)$,
\bea
F_{ij} (W^\pm) & = & \frac{1}{16 \pi^2} \cdot e \cdot g m_W \cdot g c_W \cdot \frac{1}{m_W^2} \cdot {\mathcal O}^S_{1i} {\mathcal O}^G_{1j} \nonumber \\
& \times & 
\left\{
\left[ 5 + \frac{2}{ \tau_{iW}} + \left( 1 + \frac{2}{ \tau_{iW}} \right) {\left( 1 -  \frac{4}{ \lambda_{jW}} \right) } \right] I_1 \left( \tau_{iW}, \lambda_{jW} \right) \right. \nonumber \\
&  &  \left.
- 16 {\left( 1 -  \frac{1}{ \lambda_{jW}} \right)} I_2 \left( \tau_{iW}, \lambda_{jW} \right) \right\} \; .
\label{FWloopZg}
\eea
{We note that the two factors $(1 - 4/\lambda_{jW})$ and $(1 - 1/\lambda_{jW})$ in (\ref{FWloopZg}) were not taken into account 
properly in~\cite{Huang:2015wts}!}
Here and in the following, we denote $\tau_{il} = 4 m^2_l / m^2_{h_i}$ and $\lambda_{jl} = 4 m^2_l / m^2_{Z_j}$. 
The two functions $I_1(\tau,\lambda)$ and $I_2(\tau,\lambda)$ are well known and given by~\cite{Gunion:1989we}
\bea
 \label{I1}
I_1(\tau,\lambda) & = & \frac{\tau \lambda}{2 ( \tau - \lambda)} + \frac{\tau^2 \lambda^2}{2 ( \tau - \lambda )^2} \left[ f(\tau) - f(\lambda) \right]
 + \frac{\tau^2 \lambda}{2 ( \tau - \lambda )^2} \left[ g(\tau) - g(\lambda) \right] \; , \\
I_2(\tau,\lambda) & = & - \frac{\tau \lambda}{2 ( \tau - \lambda )} \left[ f(\tau) - f(\lambda) \right] \; ,
\label{I2}
\eea 
with 
\bea
\label{fx}
f( x ) & = & \left\{ 
\begin{array}{cr}
 \left[ {\rm arcsin} (1 / \sqrt{x} ) \right]^2  \, , &   (x \geq 1) \, , \\
- \frac{1}{4} \left[ \ln \left( \eta_+ / \eta_- \right) - i \pi \right]^2 \, ,  & (x < 1) \, ; \end{array} \right. \\
g(x) & = & \left\{ 
\begin{array}{cr}
\sqrt{x - 1} \, {\rm arcsin} (1 / \sqrt{x} ) \, , &   (x \geq 1) \, , \\
\frac{1}{2} \sqrt{1 - x} \left[ \ln \left( \eta_+ / \eta_- \right) - i \pi \right] \, ,  & (x < 1) \, ; \end{array} \right.
\label{gx}
\eea
where 
\be
\eta_\pm \equiv 1 \pm \sqrt{1 - x} \; .
\ee
We note that the arguments of the functions $f(x)$ and $g(x)$ in (\ref{fx}) and (\ref{gx}) are defined to be the inverse of those in~\cite{Huang:2015wts}.

All the charged fermions in G2HDM, including both the SM fermions $f^{\rm SM}$ and the new heavy fermions $f^{H}$ contribute to 
$F^{1/2}_{ij}$. Thus
\be
F^{1/2}_{i j } = \sum_{f^{\rm SM}} F_{ij} ( f^{\rm SM} ) + \sum_{f^H} F_{ij}  ( f^H ) \; ,
\label{Fonehalf}
\ee
where
\bea
F_{ij } ( f^{\rm SM} ) & = &  \frac{1}{16 \pi^2} \cdot N^c_{ f^{\rm SM}} \cdot e Q_{ f^{\rm SM}}  \cdot \frac{m_{ f^{\rm SM}}} {v} 
 \cdot C^{f^{\rm SM}}_{Vj} \cdot \frac{-2}{m_{f^{\rm SM}}} \cdot {\mathcal O}^S_{1i} \nonumber \\
 & \, & \times \biggl[ I_1 \left( \tau_{if^{\rm SM}} , \lambda_{jf^{\rm SM}}  \right) - I_2 \left( \tau_{if^{\rm SM}} , \lambda_{jf^{\rm SM}}  
 \right) \biggr] \; ,
 \label{FfSMloopZg}
\eea
and
\bea
F_{ij} ( f^{\rm H} ) & = &  \frac{1}{16 \pi^2} \cdot N^c_{ f^{\rm H}} \cdot e Q_{ f^{\rm H}}  \cdot \frac{m_{ f^{\rm H}}} {v_\Phi} 
 \cdot C^{f^{\rm H}}_{Vj} \cdot \frac{-2}{m_{f^{\rm H}}} \cdot {{\mathcal O}^S_{2i} } \nonumber \\
 & \, & \times \biggl[ I_1 \left( \tau_{if^{\rm H}} , \lambda_{jf^{\rm H}}  \right) - I_2 \left( \tau_{if^{\rm H}} , \lambda_{jf^{\rm H}}  
 \right) \biggr] \; ,
 \label{FfHloopZg}
\eea
with $N^c_{f}$ being the color factor and $Q_f$ the electric charge of $f$ in unit of $e > 0$;
$C_{Vj}^f$ is the vector coupling of $Z_j$ with fermion $f$ given by~\footnote{
It is well known that the axial vector couplings  
$C^f_{Aj} =  ( - C^f_{Lj} + C^f_{Rj} ) / 2$ 
do not contribute to the $h_i \to Z_j\gamma$ fermion loop amplitudes.}
\be
C^f_{Vj} = \frac{1}{2} \left( C^f_{Lj} + C^f_{Rj} \right) \; ,
\label{CfVj}
\ee
where
\begin{table}[htbp!]
\begin{tabular}{l|l}
\hline
$C^u_{Lj}$ & $ \frac{g}{c_W} \left( \frac{1}{2} - \frac{2}{3} s^2_W \right)  {\mathcal O}^G_{1j} $  \\
$C^u_{Rj}$ & $ - \frac{g}{c_W} \left(  \frac{2}{3} \right) s^2_W {\mathcal O}^G_{1j}  + g_H  \left( + \frac{1}{2} \right) {\mathcal O}^G_{2j} + {\frac{1}{2}}  g_X  {\mathcal O}^G_{3j}$ \\
$C^d_{Lj}$ &  $ \frac{g}{c_W} \left( - \frac{1}{2} + \frac{1}{3} s^2_W \right)  {\mathcal O}^G_{1j} $\\
$C^d_{Rj}$ & $ - \frac{g}{c_W} \left( - \frac{1}{3} \right) s^2_W {\mathcal O}^G_{1j}  + g_H  \left( - \frac{1}{2} \right) {\mathcal O}^G_{2j} +   g_X \left( - {\frac{1}{2}}  \right) {\mathcal O}^G_{3j} $ \\
$C^{u^H}_{Lj}$ & $ - \frac{g}{c_W} \left( \frac{2}{3} \right) s^2_W {\mathcal O}^G_{1j} $ \\
$C^{u^H}_{Rj}$ &  $ - \frac{g}{c_W} \left(  \frac{2}{3} \right) s^2_W {\mathcal O}^G_{1j}  + g_H  \left( - \frac{1}{2} \right) {\mathcal O}^G_{2j} +  {\frac{1}{2}}  g_X {\mathcal O}^G_{3j} $ \\
$C^{d^H}_{Lj}$ &  $ - \frac{g}{c_W} \left( - \frac{1}{3} \right) s^2_W {\mathcal O}^G_{1j} $ \\
$C^{d^H}_{Rj}$ &  $ - \frac{g}{c_W} \left( - \frac{1}{3} \right) s^2_W {\mathcal O}^G_{1j}  + g_H  \left( + \frac{1}{2} \right) {\mathcal O}^G_{2j} +   g_X \left( - {\frac{1}{2}}  \right) {\mathcal O}^G_{3j}$ \\
\hline
\end{tabular}
\caption{Coupling coefficients $C^f_{Lj}$ and $C^f_{Rj}$ for quarks.}
\label{tab:CCQ}
\end{table}

\begin{table}[htbp!]
\begin{tabular}{l|l}
\hline
$C^\nu_{Lj}$ & $ \frac{g}{c_W} \left( + \frac{1}{2} \right) {\mathcal O}^G_{1j} $ \\
$C^\nu_{Rj}$ & $  g_H  \left( + \frac{1}{2} \right) {\mathcal O}^G_{2j} +  {\frac{1}{2}}  g_X  {\mathcal O}^G_{3j} $ \\
$C^e_{Lj}$ &   $ \frac{g}{c_W} \left( - \frac{1}{2} + s^2_W \right) {\mathcal O}^G_{1j} $ \\
$C^e_{Rj}$ &  $ - \frac{g}{c_W} \left( - 1 \right) s^2_W {\mathcal O}^G_{1j}  + g_H  \left( - \frac{1}{2} \right) {\mathcal O}^G_{2j} +   g_X \left( -  {\frac{1}{2}} \right) {\mathcal O}^G_{3j} $ \\
$C^{\nu^H}_{Lj}$ &  0 \\
$C^{\nu^H}_{Rj}$ &   $  g_H  \left( - \frac{1}{2} \right) {\mathcal O}^G_{2j} +  {\frac{1}{2}}  g_X  {\mathcal O}^G_{3j} $ \\
$C^{e^H}_{Lj}$ &   $ - \frac{g}{c_W} \left( - 1 \right) s^2_W {\mathcal O}^G_{1j} $ \\
$C^{e^H}_{Rj}$ &  $ - \frac{g}{c_W} \left( - 1 \right) s^2_W {\mathcal O}^G_{1j}  + g_H  \left( + \frac{1}{2} \right) {\mathcal O}^G_{2j} +   g_X \left( - {\frac{1}{2}}  \right) {\mathcal O}^G_{3j} $ \\
\hline
\end{tabular}
\caption{Coupling coefficients $C^f_{Lj}$ and $C^f_{Rj}$ for leptons.
}
\label{tab:CCL}
\end{table}
\bea
\label{CfSMLj}
C^{f^{\rm SM}}_{Lj} & = & \frac{g}{c_W} \left( T^3_L \left( f^{\rm SM} \right) - Q  \left( f^{\rm SM} \right) s^2_W \right) {\mathcal O}^G_{1j} \; ,\\
\label{CfSMRj}
C^{f^{\rm SM}}_{Rj} & = & \frac{g}{c_W} \left( - Q  \left( f^{\rm SM} \right) s^2_W \right) {\mathcal O}^G_{1j} 
+ g_H T^3_H \left( f^{\rm SM} \right) {\mathcal O}^G_{2j} + g_X Q_X \left( f^{\rm SM} \right) {\mathcal O}^G_{3j} \; , \\
\label{CfHLj}
C^{f^{H}}_{Lj} & = & \frac{g}{c_W} \left( - Q  \left( f^{H} \right) s^2_W \right) {\mathcal O}^G_{1j} \; , \\
C^{f^{H}}_{Rj} & = & \frac{g}{c_W} \left( - Q  \left( f^{H} \right) s^2_W \right) {\mathcal O}^G_{1j} 
+ g_H T^3_H \left( f^{H} \right) {\mathcal O}^G_{2j} + g_X Q_X \left( f^{H} \right) {\mathcal O}^G_{3j}  \; .
\label{CfHRj}
\eea
$T^3_{L,H}$ is the third component of the generators of $SU(2)_{L,H}$, $Q=T^3_L + Y$ is the electric charge, and $Q_X$ is the $U(1)_X$ charge.
The explicit expressions for  $C^f_{Lj}$ and $ C^f_{Rj}$ for both the SM fermions and new heavy fermions in G2HDM are listed in Table~\ref{tab:CCQ} 
for quarks and Table~\ref{tab:CCL} for leptons. The vector couplings $C^f_{Vj}$ of quarks and leptons are listed in Table~\ref{tab:CCQV} 
and Table~\ref{tab:CCLV} respectively.

\begin{table}[htbp!]
\begin{tabular}{l|l}
\hline
$C^u_{Vj}$ & $ \frac{1}{2} \left[ \frac{g}{c_W} \left( \frac{1}{2} - \frac{4}{3} s^2_W \right)  {\mathcal O}^G_{1j} 
+ g_H  \left( + \frac{1}{2} \right) {\mathcal O}^G_{2j} +  {\frac{1}{2}}  g_X  {\mathcal O}^G_{3j} \right] $  \\
$C^d_{Vj}$ &  $ \frac{1}{2} \left[ \frac{g}{c_W} \left( - \frac{1}{2} + \frac{2}{3} s^2_W \right)  {\mathcal O}^G_{1j} 
+ g_H  \left( - \frac{1}{2} \right) {\mathcal O}^G_{2j} +    g_X \left( - {\frac{1}{2}} \right) {\mathcal O}^G_{3j} \right] $\\
$C^{u^H}_{Vj}$ &  $ \frac{1}{2} \left[ - \frac{g}{c_W} \left(  \frac{4}{3} \right) s^2_W {\mathcal O}^G_{1j}  + g_H  \left( - \frac{1}{2} \right) {\mathcal O}^G_{2j} +  {\frac{1}{2}}  g_X {\mathcal O}^G_{3j} \right] $ \\
$C^{d^H}_{Vj}$ &  $ \frac{1}{2} \left[  \frac{g}{c_W} \left( \frac{2}{3} \right) s^2_W {\mathcal O}^G_{1j}  + g_H  \left( + \frac{1}{2} \right) {\mathcal O}^G_{2j} +   g_X \left( - {\frac{1}{2}}  \right) {\mathcal O}^G_{3j} \right] $ \\
\hline
\end{tabular}
\caption{Coupling coefficients $C^f_{Vj}$ for quarks.}
\label{tab:CCQV}
\end{table}
\begin{table}[htbp!]
\begin{tabular}{l|l}
\hline
$C^\nu_{Vj}$ & $ \frac{1}{2} \left[ \frac{g}{c_W} \left( + \frac{1}{2} \right) {\mathcal O}^G_{1j} 
+ g_H  \left( + \frac{1}{2} \right) {\mathcal O}^G_{2j} +  {\frac{1}{2}}  g_X  {\mathcal O}^G_{3j} \right] $ \\
$C^e_{Vj}$ &   $  \frac{1}{2} \left[ \frac{g}{c_W} \left( - \frac{1}{2} + 2 s^2_W \right) {\mathcal O}^G_{1j}  
+ g_H  \left( - \frac{1}{2} \right) {\mathcal O}^G_{2j} +    g_X \left( - {\frac{1}{2}} \right) {\mathcal O}^G_{3j} \right] $ \\
$C^{\nu^H}_{Vj}$ &   $  \frac{1}{2} \left[ g_H  \left( - \frac{1}{2} \right) {\mathcal O}^G_{2j} + {\frac{1}{2}}   g_X  {\mathcal O}^G_{3j} \right] $ \\
$C^{e^H}_{Vj}$ &  $ \frac{1}{2} \left[ \frac{g}{c_W} \left( 2 \right) s^2_W {\mathcal O}^G_{1j}  + g_H  \left( + \frac{1}{2} \right) {\mathcal O}^G_{2j} +   g_X \left( -{\frac{1}{2}}  \right) {\mathcal O}^G_{3j}  \right] $ \\
\hline
\end{tabular}
\caption{Coupling coefficients $C^f_{Vj}$ for leptons.
}
\label{tab:CCLV}
\end{table}

There is only one charged Higgs $H^\pm$ in G2HDM. Thus $F^0_{ij} = F_{ij} ( H^\pm )$ with
\be
F_{ij} (H^\pm) =  \frac{1}{16 \pi^2} \cdot e Q_{H^+} \cdot g_{h_i H^+H^-} \cdot g_{Z_j H^+ H^-} \cdot \frac{2}{m^2_{H^\pm}} \cdot 
I_1 \left( \tau_{i H^\pm} , \lambda_{j H^\pm}  \right) \; ,
\label{FCHloopZg}
\ee
where $Q_{H^+}  = +1$, and $g_{h_i H^+H^-}$ and $g_{Z_j H^+ H^-}$ are the $h_i H^+ H^-$ and $Z_j H^+ H^-$ couplings in the G2HDM 
respectively. Explicitly they are
\bea
\label{ghiCHCH}
g_{h_i H^+H^-} & = & \left( 2 \lambda_H - \lambda^\prime_H \right) v {\mathcal O}^S_{1i} 
+ \left( \lambda_{H \Phi} + \lambda^\prime_{H \Phi} \right) v_\Phi \mathcal{O}^S_{2i} \; , \\
g_{Z_j H^+H^-} & = &  \; \frac{1}{2} ( g \, c_W - g^\prime s_W ) {\mathcal O}^G_{1j} - \frac{1}{2} g_H {\mathcal O}^G_{2j} +{\frac{1}{2} } g_X {\mathcal O}^G_{3j} \; .
\label{gZjCHCH}
\eea

\newpage 

\noindent
\underline{$h_i \to \gamma \gamma$}

The decay rate for $h_i \to \gamma  \gamma $ is 
\be
\Gamma( h_i \to \gamma \gamma ) = \frac{1}{64 \pi} m_{h_i}^3 \left( 1 - \frac{m_{Z_j}^2}{m_{h_i}^2} \right)^3 
\biggl\vert F^1_{ i }  + F^{1/2}_{ i }  + F^0_{ i }  \biggr\vert^2 \; ,
\label{Ratehigg}
\ee
with similar definitions of $F^s_{i}$ like $F^s_{ij}$ before. The factor of $1/64$ instead of $1/32$ is due to Bose statistics in the
diphoton final state. The previous formulas of $F_{ij}(W^\pm)$, $F_{ij}( f^{\rm SM} )$, $F_{ij}( f^H )$ and $F_{ij}( H^\pm )$ 
can be easily translated into the diphoton case by taking $m_{Z_j} \to 0$ limit or equivalently $\lambda_{jl} = 4 m^2_l/m^2_{Z_j} \to \infty$ and noting that
\bea
\label{I1limit}
I_1 ( \tau, \infty ) & = & - \frac{\tau}{2} \left( 1 - \tau f (\tau ) \right) \; ,  \\
I_2 (\tau, \infty ) & = & \frac{\tau}{2} f ( \tau ) \; .
\label{I2limit}
\eea
The results are
$F^1_{ i } = F_{ i }( W^\pm )$, 
$F^{1/2}_{ i } = \sum_{f^{\rm SM}} F_{ i }( f^{\rm SM} ) + 
\sum_{f^H} F_{ i }( f^H ) $, and
$F^0_{ i } = F_{ i }( H^\pm )$ with
\bea
\label{FWloopgg}
F_{ i } ( W^\pm ) & = & 
\frac{1}{16 \pi^2} \cdot e^2 \cdot g m_W \cdot \frac{-1}{m_W^2} \cdot {\mathcal O}^S_{1i} \cdot
\left[ 2 + 3 \tau_{iW} + 3 \tau_{iW} \left( 2 -  \tau_{iW} \right) f (  \tau_{iW} ) \right]
\, ,  \\
\label{FfSMloopgg}
F_{ i } ( f^{\rm SM} ) & = & 
 \frac{1}{16 \pi^2} \cdot N^c_{ f^{\rm SM}} \cdot e^2 Q^2_{ f^{\rm SM}}  \cdot \frac{m_{ f^{\rm SM}}} {v} 
\cdot \frac{4}{m_{f^{\rm SM}}} \cdot {\mathcal O}^S_{1i} \nonumber \\
& & \times \left\{ \tau_{if^{\rm SM}} \left[ 1 + \left( 1 - \tau_{if^{\rm SM}} \right) f ( \tau_{if^{\rm SM}} ) \right] \right\}
\; ,  \\
\label{FfHloopgg}
F_{ i } ( f^H ) & = & 
 \frac{1}{16 \pi^2} \cdot N^c_{ f^{\rm H}} \cdot e^2 Q^2_{ f^{\rm H}}  \cdot \frac{m_{ f^{\rm H}}} {v_\Phi} 
\cdot \frac{4}{m_{f^{\rm H}}} \cdot { {\mathcal O}^S_{2i} } \nonumber \\
& & \times \left\{ \tau_{if^{\rm H}} \left[ 1 + \left( 1 - \tau_{if^{\rm H}} \right) f ( \tau_{if^{\rm H}} ) \right] \right\}
\; ,  \\
\label{FCHloopgg}
F_{ i } ( H^\pm ) & = & \frac{1}{16 \pi^2} \cdot e^2 \cdot g_{h_i H^+ H^-} \cdot \frac{-1 }{ m^2_{H^\pm} } \cdot 
\left\{ \tau_{iH^\pm} \left[ 1 -  \tau_{iH^\pm} f( \tau_{iH^\pm}) \right] \right\}  \; .
\eea
Note that both the charged Higgs and new heavy fermion contributions were not handled correctly 
in $h_i \to \gamma\gamma$ and  $h_i \to Z_j\gamma$ in~\cite{Huang:2015wts}.


\end{document}